\documentstyle[epsf,eqsecnum,prc,preprint,aps]{revtex}
\begin{document}
\tighten
\draft
\def\be{\begin{equation}}
\def\ee{\end{equation}}
\def\bea{\begin{eqnarray}}
\def\eea{\end{eqnarray}}
\def\gn{\gamma_\nu}
\def\g5{\gamma_5}
\def\vamn{\varepsilon^{a m n}}
\def\ot{(1\,\leftrightarrow\,2)}
\renewcommand{\thefootnote}{\fnsymbol{footnote}}
\title
{\bf Weak axial exchange currents for the Bethe-Salpeter equation}
\author{F.~C.~Khanna$^{\,a,\,b}$, E.~Truhl\'{\i}k$^{\,c}$}
\address{
\begin{center}
\it{
\small{
$^a$ Theoretical Physics Institute, Department of Physics, University
of Alberta, Edmonton, Alberta,Canada,T6G 2J1 \\
$^b$ TRIUMF, 4004 Wesbrook Mall, Vancouver, BC, Canada, V6T 2A3 \\
$^c$ Institute of Nuclear Physics, Czech Academy of Sciences, CZ-25068
$\check{R}$e$\check{z}$, Czech Republic
}}
\end{center}
}
\maketitle
\begin{abstract}
We construct  weak axial one-boson exchange currents for the Bethe-Salpeter equation, 
starting from chiral Lagrangians of the N$\Delta(1236)\pi \rho \,a_1 \omega$ system. 
The currents fulfil the Ward-Takahashi identities and the matrix element of the 
full current between the two-body solutions of the Bethe-Salpeter equation satisfies 
the PCAC constraint exactly.
\end{abstract}
\input feynman

\section{Introduction}
\label{intro}

The quantitative studies of the (anti)neutrino interaction with nuclei have now become 
reality, thanks to the existence of several neutrino detectors. Actually, the 
main goal of all these studies is a firm empirical evidence for the neutrino oscillations \cite{Zu},
\cite{JB}, \cite{GK}.

Here we deal with the (anti)neutrino-deuteron reactions,
\bea
\nu\,+\,d\,&\longrightarrow&\,\nu\,+\,n\,+\,p\,, \label{NCN} \\
\nu_e\,+\,d\,&\longrightarrow&\,e^-\,+\,p\,+\,p\,,  \label{CCN} \\
\overline{\nu}\,+\,d\,&\longrightarrow&\,\overline{\nu}\,+\,n\,+\,p\,, \label{NCA} \\
\overline{\nu}_e\,+\,d\,&\longrightarrow&\,e^+\,+\,n\,+\,n\,.   \label{CCA}
\eea
In the recent experiment \cite{Rl} on the electron antineutrino-induced 
deuteron disintegration
done at the Bugey reactor, the neutral current (NCA) reaction (\ref{NCA})
and the charge current (CCA)  deuteron disintegration (\ref{CCA})
were probed at low energies ($\sim$ 1 MeV). While the measured cross section for the
reaction NCA is in agreement with the previous measurements \cite{Re,Vi} and calculations
\cite{FV,TK1}, the new CCA cross section differs considerably from the previous results.

Reactions (\ref{NCN}) (NCN) and (\ref{CCN}) (CCN) are important for studying the 
solar electron neutrino oscillations. They are the main object of the 
SNO detector \cite{SNO} and should be measured soon. While the charge current effect will 
come presumably from the solar $^8$B neutrinos (E$_\nu\,<\,$15 MeV), the neutral current
effect should reflect the interaction of all the neutrino species. 
If the ratio of the observed rates
for reactions (\ref{CCN}) and (\ref{NCN}) would be measured significantly less than
the standard model prediction, then the mixing of the neutrino species would be
clearly demonstrated.

High energy neutrinos (E$_\nu\,>\,\,$50 MeV) can be expected from solar bursts and from
reactions of cosmic rays in the atmosphere (atmospheric neutrinos).

The cross sections for reactions (\ref{NCN})-(\ref{CCA}) up to energies 170 MeV are 
presented in \cite{Ku}. They were calculated within the frame of the standard
nuclear physics calculations using the nuclear wave functions obtained by solving the
Schroedinger equation with the commonly used nucleon-nucleon (NN) realistic 
potentials \cite{RP}. Besides the one-nucleon currents, static weak one-pion meson 
exchange currents (MECs) were also taken into account. 

At high energies, relativistic effects both in the 
potentials and currents (consistency of calculations) should be considered. 
In the standard non-relativistic
approach, relativistic corrections are calculated in the 1/M expansion scheme (M is
the nucleon mass). Actually, some of the components of the MECs
(the time component of the vector MEC and the space component of the axial MEC) 
exist only as relativistic corrections of order 1/M$^2$ which 
makes the calculations difficult.
On the other hand, relativistic calculations can be carried out from the
beginning within the framework of relativistic equations. This approach has recently
been applied to the electron-deuteron interaction at high energies \cite{AVOG},
where the general discussion was done for the Bethe-Salpeter (BS) \cite{BS} equation and furthermore
the electromagnetic interaction has been constructed in accord with the 
spectator (Gross) quasipotential equation. 
Here we apply the approach of the above mentioned paper to study
the structure of the weak interaction in conjunction with the BS equation, having in mind 
to perform later a similar study for the Gross equation and to use the weak MECs 
in calculations of the neutrino-deuteron cross sections at high energies.

The search for the MEC effects at low and intermediate energies
has a long and successful history \cite{KDR,ITO,KT,To,Ma,Ri}. The one-pion range
MEC effects were firmly established in the 
reactions governed by the space part of the electromagnetic current and by the 
time component of the weak axial current which are of the same order $\sim$ 1/M as the 
corresponding components of the one-nucleon current. Actually, the recent 
precise measurement \cite{Vo} of the transition rate for the reaction 
\begin{equation}
\mu^-\,+\,^{3}He\,\longrightarrow\,\nu_\mu\,+\,^{3}H   \label{mu3He}
\end{equation}
and its comparison with the calculations \cite{CT} demonstrates clearly also the presence
of the effect of the space component of the weak axial MEC.
At higher energies, other meson exchanges \cite{TS,ACo} and relativistic effects
start to play an important role.
 
At low energies, the starting point for constructing the MEC operators are 
the low-energy theorems. In order to pass to higher 
energies, one employs chiral Lagrangians relevant to a nuclear system under
investigation. In constructing the Lagrangians, symmetries play an essential role and
dynamical symmetries imbedded in the Lagrangian reflect dynamical principles
governing the system under investigation. 
Such a basic principle governing the physics of electroweak processes in nuclei at
intermediate energies is 
the vector dominance model\cite{Sa}. The vector meson fields are introduced
into Lagrangians first as massless Yang-Mills (YM) gauge fields \cite{YM} belonging
to the linear realizations of the chiral symmetry. Introduction of the
mass terms for the vector meson fields by hand makes this concept internally 
inconsistent. Also the interaction of the particles with the external fields
has to be introduced by hand since there are no other charges available which
could be related to the external fields. These defects of the
concept of the massive YM fields are later removed in the hidden local 
symmetries (HLS) approach \cite{BKY,M}.

In this approach, a given global symmetry group $G_g$
of a system Lagrangian is extended to a larger one by a local group
$H_l$ and the Higgs mechanism generates the mass terms for gauge fields
of the local group in such a way that the local symmetry is conserved.
For the chiral group $G_g\,\equiv\,[SU(2)_L \times SU(2)_R]_{g}$ and
$H_l\, \equiv\, [SU(2)_L \times SU(2)_R]_{l}$ the gauge particles are
identified \cite{BKY,M} with the $\rho$- and $a_1$ mesons.
An additional extension by the group $U(1)_l$ allows one to include
the isoscalar $\omega$ meson as well \cite{KM}. Moreover, external
gauge fields, which are related  to the electroweak interactions
of the Standard Model, are included by gauging the
global chiral symmetry group $G_g$.

Let us note that the YM approach and the HLS concept with the vector meson 
fields belonging to the linear realization of the chiral symmetry are formally 
equivalent \cite{BKY,M}.
The linear and non-linear realizations of the HLS are related by the Stueckelberg
transformation \cite{St,M}. Lagrangians written in terms of fields of any
realization are equivalent: the physical content of the model
cannot depend on the parametrization of fields.

Our starting point for constructing the weak axial MECs are chiral Lagrangians of the 
N$\Delta(1236)\pi \rho\,a_1 \omega$ system. The first one is that with the YM vector meson fields 
belonging to the linear realization of the SU(2)$\times$SU(2) chiral symmetry \cite{IT1},\cite{TSA}.
It is a minimal Lagrangian and contains no more than two derivatives of
fields in each term.  Another suitable Lagrangian constructed
already within the concept of the HLS can be found in \cite{STG}. These two 
Lagrangians differ by the vertex $\pi \rho\,a_1$ which brings into the MECs
a model dependence.
The model dependence of the electromagnetic isovector one-pion MEC upon 
this difference was studied in the backward deuteron electrodisintegration 
in \cite{STG}. It influences considerably the cross sections at energies
about 800 MeV.

The chosen Lagrangian has already been used in constructing the weak axial MEC 
operator of the one-pion range \cite{TK2} within the extended S-matrix method \cite{ATA}.
In this method, the two-nucleon transition amplitude is represented by a set of 
tree Feynman graphs of the one-boson exchange type with the external nucleon lines
on-shell. It is gauge- and Lorentz-invariant and it satisfies the PCAC equation.
It is also assumed that the nuclear states are described by the Schroedinger 
equation solved with a one-boson exchange potential.
The nuclear weak axial MEC operator is obtained after subtracting the 
first iterated Born approximation from the two-nucleon transition amplitude.
Let us stress that the first iterated Born approximation generally differs from
the positive-frequency part of the nucleon Born term entering the two-nucleon 
transition amplitude in higher order in 1/M. As mentioned above, this MEC operator 
has been employed in \cite{CT} to calculate the weak axial MEC effects in reaction
(\ref{mu3He}).

Here we construct the weak axial MEC operator for the case when the 
two-nucleon nuclear states are described by the BS relativistic equation.
In Sect.\,\ref{CH1}, we present the Lagrangians.
The transition operator is constructed in Sect.\,\ref{CH2} starting from the
YM Lagrangian and its model dependence is studied in Sect.\,\ref{CH3}.
For the construction, the same Feynman diagram technique is employed as in \cite{TK2}, 
but the nucleons are off-shell. 
In order to get the correct MEC operator, one should omit the nucleon Born terms
from the transition operator, which are already accounted for when calculating
the matrix element of the one-nucleon current between the nuclear wave functions.
We show that the full BS current satisfies the correct Ward-Takahashi identity (WTI) and that
the resulting matrix element of the MEC operator between the two-body solutions of the
BS equation satisfies the PCAC constraint exactly.

Similar program for the electromagnetic interaction has been done earlier  in Ref.\,\cite{GR}.
However, Gross and Riska use the minimal substitution in the kernel of the relativistic
equation for generating the potential exchange currents, while the non-potential currents
are derived from vertices. We derive both kinds of the currents from chiral Lagrangians.
Generally, some difference can appear, since the currents derived from Lagrangians 
might be non-minimal.  

Our results and conclusions are given in Sect.\,\ref{RC}.

The investigation made here is based on the normal Lagrangians. The anomalous Lagrangians
describe processes with the change of the internal parity of particles and they are
related to the phenomenon of chiral anomaly \cite{W1}
. In our case,
the vertices $\pi \rho \omega$ and $\rho \omega\,a_1$ would play a role. They are characterized
by the presence of the tensor $\varepsilon^{\alpha \beta \gamma \delta}$ causing the
change of the internal parity. 
The relevant Lagrangians can be found in \cite{KM},\cite{STK1}. However, the construction
of these currents is beyond the scope of this paper.

The weak axial MECs based on the Lagrangian approach have recently
been studied in other models in \cite{Dm,An}.

We use the metrics defined by $(\,+\,-\,-\,-\,)$.


\section{Chiral Lagrangians    \label{CH1}}

We construct the MEC operators from a chiral 
Lagrangian of the $N \pi \rho\, a_1 \omega$ system. Its essential part
is the Lagrangian of the subsystem $\pi \rho\, a_1$.
The conventional Lagrangians of the $\pi \rho a_{1}$ system contain
trilinear and quadrilinear derivatives of the fields. They were studied already
in sixties within the framework of the YM massless compensating fields. 
As it was shown in \cite{OZ}, a minimal
effective Lagrangian ${\cal L}^{M}_{\,\pi \rho\, a_{1}}$ of the $\pi \rho a_{1}$ system can be constructed
so that it contains no more than two field derivatives in each term. 
This restriction determines the Lagrangian completely and it describes
reasonably the related elementary processes up to 1 GeV. 
This construction was extended in \cite{IT1},\cite{TSA} to the $N \Delta \pi \rho\,a_1$
system. 

However, the approach of the massless YM fields is internally inconsistent, 
as the masses of the heavy mesons should
be put into the theory by hands and they violate the underlying symmetry.
It was realized later \cite{BKY},\cite{M} that the internal inconsistency of the
YM approach can be removed in the theory of HLS.
In this theory, the masses of the heavy mesons are generated by the Higgs mechanism
and they do not violate the symmetry any more.
The HLS approach was applied also to the construction of the
Lagrangian of the $\pi \rho\, a_1$ system \cite{BKY}, where the high energy behaviour 
of the amplitudes was corrected
by a different choice of correction Lagrangians in comparison with \cite{OZ}, thus introducing a model
dependence. However, the construction \cite{BKY} was done for the heavy mesons
on--shell and the Lagrangian cannot be used for constructing the MEC operators.
Moreover, it also provides for the anomalous magnetic moment of the $a_1$ meson $\delta_{an}$=+3,
in contradiction to its commonly accepted value $\delta_{an}=-1$ \cite{OZ}. 
Both gaps in construction have recently been removed \cite{STG}.

It has been also shown in \cite{BKY},\cite{M} that the Lagrangians of the YM approach
are formally equivalent to those constructed from the fields of the linear realizations 
of the HLS and that they are a sub--manifold of the manifold of the BKY Lagrangians.
We first present the $N \Delta \pi \rho\,a_1$ Lagrangian ${\cal L}^{M}$ of the YM type 
\cite{IT1},\cite{TSA},\cite{OZ}. 
It can be easily extended to contain
also the $\omega$ meson field. Then we introduce the HLS Lagrangian ${\cal L}^{H}$
\cite{STG},\cite{STK2} and discuss  possible source of the model dependence of our MEC operators.

The YM Lagrangian  of the $N\Delta\pi\rho a_{1}$ system can be written with an arbitrary mixing of the
pseudoscalar and pseudovector $\pi NN$ couplings \cite{TSA} but we restrict ourselves to the pseudoscalar
one only. With the $\omega$ meson included, it is of the form 
\begin{equation}
{\cal L}^{M}\,=\,{\cal L}^{M}_{\,N\pi \rho\, a_{1} \omega}\,+
\,{\cal L}^{M}_{\,N\Delta\pi\rho\, a_{1}}\,+\,{\cal L}^{M}_{\,\pi \rho\, a_{1}}\,,  \label{L}
\end{equation}
\begin{eqnarray}
{\cal L}^{M}_{\,N\pi \rho\, a_{1}\omega}\,& = & {\cal L}^{M\,(0)}_{\,N\pi \rho\, a_{1}\omega} 
+\frac{g_{\rho}}{2f_{\pi}}\,(1-2 g_{\,A}^{\,2})
\,\bar{\Psi}\gamma^{\mu}(\vec{\tau}\,\cdot\,\vec{\pi}\,\times\,
\vec{a}_{\mu})\Psi   \nonumber \\
& & +\frac{g_{\rho}}{2}\bar{\Psi}
\left (\,\gamma^{\mu}\vec{\rho}_{\mu}+
\frac{\kappa^{V}_{\rho}}{4M}\,\sigma^{\mu\nu}
\vec{\stackrel{\sim}{\rho}}^{\,\prime}_{\mu\nu}\right)\,\cdot
\,\vec{\tau}\Psi 
+\frac{i}{4} g_{\rho} \frac{g}{M}
\frac{\kappa^{V}_{\rho}}{2M}\bar{\Psi}\gamma_{5}\sigma^{\mu\nu}
\left (\,\vec{\stackrel{\sim}{\rho}}^{\,\prime}_{\mu\nu}\,\cdot\,\vec{\pi}
\,\right )\Psi\,  \nonumber   \\
& & +\,{\cal O} (\pi^{2})\,,   \label{LNPRA1O}
\end{eqnarray}
\begin{eqnarray}
{\cal L}^{M\,(0)}_{\,N\pi \rho\, a_{1}\omega}\,& = & \bar{\Psi}\left(i\not{\partial}
\,-\,M\right)\Psi\,-\,ig\bar{\Psi}\,\vec{\pi}\,
\cdot\,\vec{\tau}\gamma_{5}\Psi  
 -g_{A}g_{\rho}\bar{\Psi}\gamma^{\mu}\gamma_{5}
(\vec{\tau}\cdot\vec{a}_{\mu})\Psi   \nonumber \\
& & +\frac{M}{2}\left (\frac{g_A}{f_\pi}\right )^{\,2}\,\bar{\Psi}
\,\vec{\pi}^{\,2}\,\Psi  +\,\left (\frac{g_{A}}
{2f_{\pi}}\right )^{\,2}\,\bar{\Psi}
\gamma^{\mu}(\vec{\tau}\,\cdot\,\vec{\pi}\,\times\,\partial_\mu\vec{\pi})\Psi  \nonumber  \\
& & +\frac{g_{\omega}}{2}\bar{\Psi}
\left (\,\gamma^{\mu}\omega_{\mu}+
\frac{\kappa^{S}}{4M}\,\sigma^{\mu\nu}
\omega_{\mu\nu}\right)\,\Psi\,  
+\frac{i}{4} g_{\omega} \frac{g}{M}
\frac{\kappa^{S}}{2M}\bar{\Psi}\gamma_{5}\sigma^{\mu\nu}
\left (\,\vec{\pi}\cdot\vec{\tau}\right)\omega_{\mu\nu}\,\Psi\, \nonumber  \\
& & +\,{\cal O} (\pi^{2})\,.   \label{LNPRA1O0}
\end{eqnarray}

Here we define
\begin{eqnarray*}
\vec{\stackrel{\sim}{\rho}}^{\,\prime}_{\mu\nu}=\vec{\rho}_
{\mu\nu}+g_{\rho}\vec{\rho}_{\mu}\times\vec{\rho}_{\nu}+
\frac{1}{f_{\pi}}[\vec{a}_{\mu}\times\partial_{\nu}\vec{\pi}
-\vec{a}_{\nu}\times\partial_{\mu}\vec{\pi}
+\vec{\pi}\times\vec{a}_{\mu\nu}\,]+{\cal O} (\pi^{2})\,,
\end{eqnarray*}
and
\begin{eqnarray*}
\vec{\rho}_{\mu\nu}\,=\,\partial_{\mu}\vec{\rho}_{\nu}\,-\,
\partial_{\nu}\vec{\rho}_{\mu}
\,,\qquad
\vec{a}_{\mu\nu}\,=\,\partial_{\mu}\vec{a}_{\nu}\,-\,
\partial_{\nu}\vec{a}_{\mu}\,,\qquad
\omega_{\mu \nu}\,=\,\partial_\mu \omega_\nu\,-\,\partial_\nu \omega_\mu\:.
\end{eqnarray*}
Furthermore
\be
{\cal L}^{M}_{\,N\Delta\pi\rho a_{1}}\,  = \, 2\frac{f_{\pi N \Delta}}
{m_{\pi}}\,\bar{\Psi}^{\,\mu}\,\vec{T}\,\Psi\,\cdot\,\nabla_{\,\mu}
\vec{\pi}\,+i\,g_{\rho}\frac{G_{1}}{M}\,\bar{\Psi}^{\,\mu}\,
\gamma_{5}\gamma^{\nu}\vec{T}\,\Psi\,\cdot\,
\vec{\rho}_{\mu\nu}\,+\,h.\,c.   \label{LNDPRA1}
\ee
Here the transition isospin operator $\vec{T}$
is defined \cite{BW} as
\begin{equation}
\left(\,\vec{T}\,\right)_{\,M_{T}\,m_{t}}\,=\,\sum_{l}
\,\left(\,\frac{3}{2}\,M_{T}\,\Big|\,1\,l\,\frac{1}{2}\,m_{t}\,
\right)\,\vec{t}^{\:*\,l}\:,
\qquad
\vec{t}^{\:\,\pm}\,=\,\mp\,\frac{1}{\sqrt{2}}\,\left( \begin{array}{c}
1 \\ \pm\,i \\ 0 \end{array} \right)\,,\quad
\vec{t}^{\:\,0}\,=\,\left( \begin{array}{c} 0 \\ 0 \\ 1 \end{array} \right)\,.  \label{TS}
\end{equation}
Further in Eq.\,(\ref{LNDPRA1})
\begin{equation}
\nabla_{\,\mu}\,\vec{\pi}\,=\,\frac{1}{2}\,\partial_{\mu}\vec{\pi}\,
-\,f_{\pi}g_{A_{1}}\vec{a}_{\mu}\,+\,{\cal O} (\pi^{2})\,.
\end{equation}

As it was discussed in \cite{IT1}, $G_{1}\approx 2.6$.
Actually, another term $\sim\,G_2$ present in the original Lagrangian
was neglected in (\ref{LNDPRA1})  because of one additional factor $\sim\,1/M_\Delta$.

The Lagrangian of the $\pi \rho a_{1}$ system is of the form
\begin{eqnarray}
{\cal L}^{M}_{\pi \rho a_{1}}\, & = & \,g_{\rho}\vec{\rho}^{\,\mu}\,\cdot\,
\vec{\pi}\times\partial_{\mu}\vec{\pi}\,+\,g_{\rho}\vec{\rho}^{\,\mu}
\times\vec{\rho}^{\,\nu}\,\cdot\,\partial_{\nu}\vec{\rho}_{\mu}\,
-\,g_{\rho}(\vec{\rho}^{\,\mu}\times\vec{a}^{\,\nu} \nonumber \\
& & -\vec{\rho}^{\,\nu}\times\vec{a}^{\,\mu})\,\cdot\,
\partial_{\mu}\vec{a}_{\nu}\,-\,\frac{1}{f_{\pi}}\,
\vec{\rho}^{\,\mu\nu}\,\cdot\,(\vec{a}_{\mu}\times\partial_{\nu}
\vec{\pi}\,+\,\frac{1}{4}\vec{\pi}\times\vec{a}_{\mu\nu})\,.   \label{LPRA1}
\end{eqnarray}

In this approach, the hadron currents are mediated by mesons,
\begin{equation}
{J}^{S\,\mu}_{M}\,=\,-\frac{m^{2}_{\omega}}{g_{\omega}}\,{\omega}^{\,\mu}\:,
\vec{J}^{\,\mu}_{M}\,=\,-\frac{m^{2}_{\rho}}{g_{\rho}}\,\vec{\rho}^{\,\mu}\:,
\vec{J}^{\,5\,\mu}_{M}\,=\,-
\frac{m^{2}_{\rho}}{g_{\rho}}\,\vec{a}^{\,\mu}\,
-\,f_{\pi}\partial^{\mu}\vec{\pi}\,-\,f_{\pi}g_{\rho}
\vec{\rho}^{\,\mu}\times\vec{\pi}\,+\,{\cal O} (\pi^{2})\,.   \label{OBYMCS}
\end{equation}

The Lagrangian ${\cal L}^{M}$ (\ref{L}) together with the currents
(\ref{OBYMCS}) consist of all the vertices necessary for constructing the
operators of the isovector vector and axial one--nucleon and
$\pi$, $\rho$, $a_{1}$ and $\omega$ MECs in the tree approximation.

The HLS Lagrangian ${\cal L}^{H}$ consists of three terms like the Lagrangian
${\cal L}^{M}$, but now the heavy meson fields represent the nonlinear 
realization of the HLS for the groups $H_l\,\equiv\,[SU(2)_L \times SU(2)_R]_l
\times U(1)_l$. In comparison with (\ref{LNPRA1O}) we have
\begin{eqnarray}
{\cal L}^{H}_{\,N\pi \rho\, a_{1}\omega}\,& = & {\cal L}^{H\,(0)}_{\,N\pi \rho\, a_{1}\omega}
-g_\rho\frac{g_A}{2f_\pi}\,\bar{\Psi}
\,\gamma^\mu \g5\,(\vec \tau \cdot \vec \rho_\mu \times \vec \Pi )\,\Psi  
+g_\rho \frac{g^2_A}{f_{\pi}}
\,\bar{\Psi}\gamma^{\mu}(\vec{\tau}\,\cdot\,\vec a_\mu \times \vec{\Pi} )\Psi   \nonumber \\
& & +\frac{g_{\rho}}{2}\bar{\Psi}
\left (\,\gamma^{\mu}\vec{\rho}_{\mu}+
\frac{\kappa^{V}_{\rho}}{4M}\,\sigma^{\mu\nu}
\vec{\cal F}^{(\rho)}_{\mu\nu}\right)\,\cdot
\,\vec{\tau}\Psi 
+\frac{i}{4} g_{\rho} \frac{g}{M}
\frac{\kappa^{V}_{\rho}}{2M}\bar{\Psi}\gamma_{5}\sigma^{\mu\nu}
\left (\,\vec{\cal F}^{(\rho)}_{\mu\nu}\,\cdot\,\vec{\Pi}
\,\right )\Psi\,  \nonumber   \\
& & +\,{\cal O} (\pi^{2})\,,   \label{LhlsNPRA1O}
\end{eqnarray}
with 
\be
\vec{\cal F}^{(\rho)}_{\mu\nu}\,=\,\vec{\rho}_
{\mu\nu}+g_{\rho}\vec{\rho}_{\mu}\times\vec{\rho}_{\nu}\,,  \label{Fr}
\ee
and the Lagrangian ${\cal L}^{H\,(0)}_{\,N\pi \rho\, a_{1}\omega}$ formally coincides
with ${\cal L}^{M\,(0)}_{\,N\pi \rho\, a_{1}\omega}$. The same holds also for
${\cal L}^{H}_{\,N\Delta\pi\rho a_{1}}$. As to the Lagrangian ${\cal L}^{H}_{\pi \rho a_{1}}$,
it is of the same form as given in Eq.\,(\ref{LPRA1}) with the last two terms 
changed to
\be
\Delta{\cal L}^{H}_{\pi \rho a_{1}}\,=\,\frac{1}{f_\pi}\,\left( \vec \rho^{\,\mu}_{\,\,\nu} 
\cdot \vec a_\mu \times \partial^\nu \vec \Pi\,+\, \frac{1}{2}\vec \rho_\mu \cdot
\partial^\nu \vec \Pi \times \vec a^{\,\mu}_{\,\,\nu} \right)\,.  \label{DLHLSPRA1}
\ee
The currents-analogues of Eqs.\,(\ref{OBYMCS}) are given by the following set of equations
\bea
{J}^{S\,\mu}_{H}\,&=&\,-\frac{m^{2}_{\omega}}{g_{\omega}}\,{\omega}^{\,\mu}\,,\quad
\vec{J}^{\,\mu}_{H}\,=\,-\frac{m^{2}_{\rho}}{g_{\rho}}\,\vec{\rho}^{\,\mu}
\,-\,2f_{\pi}g_{\rho}\vec{a}^{\,\mu}\times\vec{\Pi}\,+\,{\cal O} (\Pi^{2})\,,  \label{JSVHLS} \\
\vec{J}^{\,5\,\mu}_{H}\,&=&\,-
\frac{m^{2}_{\rho}}{g_{\rho}}\,\vec{a}^{\,\mu}\,
-\,f_{\pi}\partial^{\mu}\vec{\Pi}\,-\,2f_{\pi}g_{\rho}
\vec{\rho}^{\,\mu}\times\vec{\Pi}\, \nonumber \\
& & +\,\frac{1}{g_\rho}\,\left(-\frac{1}{2f_\pi}\,\partial^\nu \vec \Pi\,
 -\,g_\rho \vec a^\nu\,+\,
e {\cal A}^\nu \right ) \times \vec \rho^{\,\mu}_{\,\,\nu}\,+\,{\cal O} (\Pi^{2})\,.  \label{JAHLS}
\eea
In contrast to the YM currents Eq.\,(\ref{OBYMCS}), the HLS curents contain also the external
weak axial field $\vec{\cal A}^\nu$. 

Checking the Lagrangians and currents for the possible source of difference between the MECs
one finds that it can arise from the $\pi \rho\, a_1$ and NN$\pi a_1$ vertices.

\section{The axial MEC operator for the Bethe-Salpeter equation \label{CH2}}

We first write the operator of the weak axial current for the i{\it th} nucleon 
(i$\,=\,1,\,2$)

\be
\hat{J}^{a\,\mu}(1,i)\,=\,\frac{g_A}{2}\,m^2_{a_1}\,\Delta^{\mu\,\nu}_{a_1}(q)\,
\hat \Gamma^{5\,a}_{i\,\nu}\,
+\,if_\pi\,\Delta^\pi_F(q^2)\,q^\mu\,\hat \Gamma^a_i. \label{ONAC}
\ee

Here g$_A$ is the weak axial coupling (the nucleon axial charge), g$_A\,=\,$-1.256,
q$_{\,i}\,=\,$p$'{\,_i}\,-\,$p$_{\,i}$, the
vector-meson propagator is generally designed as 
\be
\Delta^{\mu\,\nu}_B(q)\,=\,\left(g^{\mu\,\nu}\,-\,\frac{q^\mu\,q^\nu}{m^2_B}\right)\,\Delta_B(q^2),
\quad\Delta_B(q^2)\,=\,\frac{1}{m^2_B\,-\,q^2}\,,  \label{prop}
\ee
and the pseudovector and pseudoscalar vertices are defined as
\be
\hat \Gamma^{5\,a}_{i\,\nu}\,=\,\left( \gn \g5 \tau^a \right )_i\,,\quad 
\hat \Gamma^a_i\,=\,ig\left (\g5 \tau^a \right )_i\,.  \label{vpvps}
\ee
The divergence of the current (\ref{ONAC}) is
\bea
q_\mu \hat{J}^{a\,\mu}(1,i)\,&=&\,\left[G^{-1}(\,p'_i\,)\,\hat{e}_A(i)\,+\,
             \hat{e}_A(i)\,G^{-1}(\,p_i\,)\right]\,+\,i\,f_\pi m^2_\pi\,\Delta^\pi_F(q^2)\,
                  \hat{\Gamma}^a_i\,,  \nonumber \\
&\equiv&\,\left[\hat{e}_A(i),\,G^{-1}_i\right]_+\,+\,i\,f_\pi m^2_\pi\,\Delta^\pi_F(q^2)\,
                  \hat{\Gamma}^a_i\,. \label{DOBCi}
\eea
Here  the operator of the nucleon axial charge $\hat{e}_A$(i) and the inverse of the
nucleon propagator $G^{-1}(p)$ are defined as
\be
\hat{e}_A(i)\,=\,g_A\,\left(\g5\frac{\tau^a}{2}\right)_i\,,\quad G^{-1}(\,p\,)\,=\,\not p\,-\,M\,. \label{def1}
\ee

The general structure of the weak axial MEC operators is given in Fig.\,1. The weak axial
interaction is mediated by the meson B which is either $\pi$ or $a_1$ meson. The range of the
current is given by the meson $B_2$ which is here $\pi$, $\rho$, $a_1$ or $\omega$ meson.
In the graphs 1a and 1b, $\cal N$ is either for the nucleon N (in this case the graphs are the 
standard nucleon Born terms) or for the $\Delta(1236)$ isobar. We shall write the corresponding currents as 
${\hat J}^{a\,\mu}_{B_2}({\cal N},\,B)$.In the case of the BS equation,
the nucleon Born terms do not enter the nuclear MECs because they are fully taken into account
when calculating the matrix element of the one-nucleon current with the solutions of 
this equation.  The graph 1c represents one type of contact terms
which we shall denote as ${\hat J}^{a\,\mu}_{c\,B_2}(B)$. It is connected with the weak
production amplitude of the $B_2$ meson on the nucleon. Another type of the contact
terms is given by the graph 1d where the weak axial current interacts directly
with the mesons $B_1$ and $B_2$.  We shall write it as ${\hat J}^{a\,\mu}_{B_1\,B_2}$.
The graph 1e is for a mesonic current which we shall write in the form
${\hat J}^{a\,\mu}_{B_1\,B_2}(B)$. 
The associated pion absorption amplitudes will be written as $\hat {\cal M}^a_{B_2}(\cal N)$, 
$\hat {\cal M}^a_{c\,B_2}$ and $\hat {\cal M}^a_{B_1\,B_2}$.
For convenience, we shall modify the notations in some cases.

\subsection{The axial MEC operator of the pion range \label{CH21}}

This operator is closely related to the one derived in \cite{TK2} from the Lagrangians
${\cal L}^{M}_{N \pi \rho\,a_1 \omega}$ and ${\cal L}^{M}_{\pi \rho\,a_1}$. Essentially, 
it has potential and non-potential parts, with the nucleon Born terms in the potential part 
omitted. Let us note that the Lagrangian used for the derivation
of the $\pi$ MEC operator corresponds to the non-linear chiral model with the pseudoscalar
$\pi$NN coupling which is connected to another model with any mixing of 
the pseudoscalar and pseudovector couplings by the unitary Foldy-Dyson transformation 
which makes all the models physically equivalent. 

In the given case, the potential amplitude consists of several contact terms where 
the weak interaction is mediated either by the a$_1$- or $\pi$ meson.
The a$_1$-pole contact terms  of the type ${\hat J}^{a\,\mu}_{c\,\pi}(a_1)$ are
\be
\hat{J}^{a\,\mu}_{c_1\,\pi}(a_1)\,=\,-\frac{1}{4f_\pi}\,m^2_{a_1}\Delta^{\mu\,\nu}_{a_1}(q)\,
\vamn \hat{\Gamma}^m_{1\,\nu}(q_1)\Delta^\pi_F(q^2_2)\,\hat{\Gamma}^n_2\,+\,\ot\,,  \label{Jnpa}
\ee
with
\be
\hat{\Gamma}^m_{i\,\nu}(q_j)\,=\,\left(\gn\,+\,i\,\frac{\kappa^V_\rho}{2M}\sigma_{\nu\,\delta}\,
q_j^{\,\delta}\right)_i\,\tau^m_i\,,    \label{Gmin}
\ee
and
\be
\hat{J}^{a\,\mu}_{c_2\,\pi}(a_1)\,=\,\frac{g^2_A}{2f_\pi}\,m^2_{a_1}\Delta^{\mu\,\nu}_{a_1}(q)\,
\vamn (\gn \tau^m)_1 \Delta^\pi_F(q^2_2)\,\hat{\Gamma}^n_2\,+\,\ot\,.  \label{JPCAC}
\ee
The $\pi$ meson contact terms of the type ${\hat J}^{a\,\mu}_{c\,\pi}(\pi)$ are of the form
\bea
\hat{J}^{a\,\mu}_{c_1\,\pi}(\pi)\,&\equiv&\,if_\pi\,q^\mu
\Delta^\pi_F(q^2)\,\hat{\cal M}^a_{c_1\,\pi}\,, \label{Jppp} \\
\Delta\hat{J}^{a\,\mu}_{c_1\,\pi}(\pi)\,&\equiv&\,if_\pi\,q^\mu
\Delta^\pi_F(q^2)\,\Delta \hat{\cal M}^a_{c_1\,\pi}\,, \label{dJppp}
\eea
where the pion absorption amplitudes are
\bea
\hat{\cal M}^a_{c_1\,\pi}\,&=&\,-\frac{i}{2}\left( \frac{g}{M} \right)^2\,
\vamn \not q_2 \tau^m_1 \Delta^\pi_F(q^2_2)\,\hat{\Gamma}^n_2\,+\,\ot\,, \label{Mappp} \\
\Delta \hat{\cal M}^a_{c_1\,\pi}\,&=&\,-\frac{i}{4}\left( \frac{g}{M} \right)^2\,
\vamn \not q_1 \tau^m_1 \Delta^\pi_F(q^2_2)\,\hat{\Gamma}^n_2\,+\,\ot\,, \label{dMappp} 
\eea
and
\bea 
\hat{J}^{a\,\mu}_{c_2\,\pi}(\pi)\,&=&\,if_\pi\,q^\mu \Delta^\pi_F(q^2)\frac{g^2}{M}
\delta_{a\,n} \Delta^\pi_F(q^2_2)\,\hat{\Gamma}^n_2\,+\,\ot\,\equiv\,
if_\pi\,q^\mu \Delta^\pi_F(q^2)\,\hat{\cal M}^a_{c_2\,\pi}\,.  \label{Jpp}
\eea

In contrast to ref.\,\cite{TK2}, besides the current (\ref{Jppp}) we have an additional
term (\ref{dJppp}) which disappears for the on-shell  nucleons.

A direct calculation of the divergence of the currents (\ref{JPCAC}), (\ref{Jppp}), (\ref{dJppp}) and
(\ref{Jpp}) yields
\bea
q_\mu \left[ \hat{J}^{a\,\mu}_{c_2\,\pi}(a_1)+\hat{J}^{a\,\mu}_{c_1\,\pi}(\pi)+
\Delta\hat{J}^{a\,\mu}_{c_1\,\pi}(\pi) 
+\hat{J}^{a\,\mu}_{c_2\,\pi}(\pi) \right]\,&=& \,if_\pi\,m^2_\pi\Delta^\pi_F(q^2)\, 
\left[ \hat{\cal M}^a_{c_1\,\pi} + 2\Delta \hat{\cal M}^a_{c_1\,\pi}  \right. \nonumber  \\
& & \hspace{-50mm} \left.+\hat{\cal M}^a_{c_2\,\pi} \right]  
-i f_\pi q^2 \Delta^\pi_F(q^2)\,\Delta \hat{\cal M}^a_{c_1\,\pi}
-if_\pi \hat{\cal M}^a_{c_2\,\pi}\,.   \label{divp1}
\eea
It can be shown that
\be
-if_\pi \hat{\cal M}^a_{c_2\,\pi}\,=\,\left[\,\hat{e}_A(1)+\hat{e}_A(2),\,\hat{V}_\pi\,\right]_+\,, \label{com}
\ee
where the one-pion exchange potential $\hat{V}_\pi$ is defined as
\be
\hat{V}_\pi\,=\,\hat{\Gamma}^n_1\,\Delta^\pi_F(q^2_2)\,\hat{\Gamma}^n_2\,.  \label{OPEP}
\ee
Taking into account Eqs.\,(\ref{dJppp}) and (\ref{com}), we can write Eq.\,(\ref{divp1}) in
the form
\bea
q_\mu \left[ \hat{J}^{a\,\mu}_{c_2\,\pi}(a_1)+\hat{J}^{a\,\mu}_{c_1\,\pi}(\pi)+
2\Delta\hat{J}^{a\,\mu}_{c_1\,\pi}(\pi) 
+\hat{J}^{a\,\mu}_{c_2\,\pi}(\pi) \right]\,&=& \,if_\pi\,m^2_\pi\Delta^\pi_F(q^2)\,
\left[ \hat{\cal M}^a_{c_1\,\pi}  \right.  \nonumber \\
& & \hspace{-70mm} \left. +2\Delta \hat{\cal M}^a_{c_1\,\pi}  +\hat{\cal M}^a_{c_2\,\pi} \right]\,+\,
\left[\,\hat{e}_A(1)+\hat{e}_A(2),\,\hat{V}_\pi\,\right]_+\,.   \label{divp2}
\eea
This continuity equation allows us to define the potential axial $\pi$ MEC operator related to the
BSE as
\be
\hat{J}^{a\,\mu}_{BS\,\pi}(p)\,=\,\hat{J}^{a\,\mu}_{c_2\,\pi}(a_1)+\hat{J}^{a\,\mu}_{c_1\,\pi}(\pi)+
2\Delta\hat{J}^{a\,\mu}_{c_1\,\pi}(\pi) +\hat{J}^{a\,\mu}_{c_2\,\pi}(\pi)\,. \label{PPIMEC}
\ee

Let us note that the current (\ref{Jnpa}) does not enter the resulting potential current (\ref{PPIMEC}).
As in the on-shell case \cite{TK2}, it satisfies the continuity equation together with the 
non-potential currents (see below) which are constructed from the $\pi \rho\, a_1$ vertices.
So we find its natural place among these non-potential currents, which we will consider now.
They are analogues of the amplitudes given in Sect.\,4 of Ref.\,\cite{TK2}. The explicit form 
of the currents is 
\bea
\hat{J}^{a\,\mu}_{\rho\,\pi}(a_1)\,&=&\,\frac{m^2_{a_1}}{8f_\pi}\,\Delta^{\,\,\mu}_{a_1\,\nu}(q)\,
\vamn \hat{\Gamma}^m_{1\,\eta}(q_1)\Delta^\rho_F(q^2_1)\left[(q_1 \cdot q_2 - q^2_1)g^{\nu\,\eta} 
+ q^\nu_1(q^\eta_1 - q^\eta_2)\right]\,
\times   \nonumber \\
& & \Delta^\pi_F(q^2_2)\,\hat{\Gamma}^n_2\,+\,\ot\,,  \label{Jpra} \\
\hat{J}^{a\,\mu}_{\rho\, \pi}(\pi)\,&=&\,if_\pi q^\mu \Delta^\pi_F(q^2)\,\left[\hat{\cal M}^a_{\rho\,\pi}
+ \Delta\hat{\cal M}^a_{\rho\,\pi} \right]\,,
\label{Jrpp} \\
\hat{\cal M}^a_{\rho\,\pi}\,&=&\,i\frac{m^2_\rho}{2f^2_\pi}\vamn q_{2\,\nu} \hat{\Gamma}_{1\,m}^{\,\nu}(q_1)
\Delta^\rho_F(q^2_1)\Delta^\pi_F(q^2_2)\,\hat{\Gamma}^n_2\,+\,\ot\,,  \label{Mrpp} \\
\Delta\hat{\cal M}^a_{\rho\,\pi}\,&=&\,i\frac{m^2_\rho}{4f^2_\pi}\vamn q_{1\,\nu} \hat{\Gamma}_{1\,m}^{\,\nu}(q_1)
\left[ 1 - 2 (q_1 \cdot q_2) \Delta^\rho_F(q^2_1)\right]
\Delta^\pi_F(q^2_2)\,\hat{\Gamma}^n_2\,+\,\ot\,,  \label{DMrpp} \\
\hat{J}^{a\,\mu}_{\rho\,\pi}\,&=&\,-\frac{m^2_\rho}{4f_\pi}
\vamn \Delta^{\mu \nu}_\rho(q_1)\hat{\Gamma}_{1\,\nu}^{\,m}(q_1)
\Delta^\rho_F(q^2_1)\Delta^\pi_F(q^2_2)\,\hat{\Gamma}^n_2\,+\,\ot\,.  \label{Jrp}
\eea

Strightforward calculation yields for the divergence of the  currents (\ref{Jnpa}) and (\ref{Jpra})-
(\ref{Jrp}) 
\be
q_\mu \left[\hat{J}^{a\,\mu}_{c_1\,\pi}(a_1)+
\hat{J}^{a\,\mu}_{\rho\,\pi}(a_1)+\hat{J}^{a\,\mu}_{\rho\, \pi}(\pi)+
\hat{J}^{a\,\mu}_{\rho\, \pi}+\Delta\hat{J}^{a\,\mu}_{\rho\,\pi}(np)\right]\,=\,if_\pi\,m^2_\pi
\Delta^\pi_F(q^2)\,\left[\hat{\cal M}^a_{\rho\, \pi}+\hat{\tilde {\cal M}}^a_{\rho\, \pi}\right]\,,
        \label{divnp1}        
\ee
where an additional current $\Delta\hat{J}^{a\,\mu}_{\rho\,\pi}(np)$ defined as
\be
\Delta\hat{J}^{a\,\mu}_{\rho\,\pi}(np)\,=\,-\frac{1}{4f_\pi}\,q^\mu \Delta^\pi_F(q^2) \vamn\,\not q_1\tau^m_1\,
\Delta^\pi_F(q^2_2)\,\hat{\Gamma}^n_2\,+\,\ot\,,  \label{dJnp}
\ee
and a new pion absorption amplitude
\be
\hat{\tilde {\cal M}}^a_{\rho\, \pi}\,=\,\frac{i}{2f_\pi^2}\, \vamn\,\not q_1\tau^m_1\,
\left[1-(q_1 \cdot q_2)\Delta^\rho_F(q^2_1)\right]\Delta^\pi_F(q^2_2)\,\hat{\Gamma}^n_2\,+\,\ot\,,  \label{Mtrp}
\ee
appear for the off-shell nucleons.

It can be verified that the sum of the currents $\hat{J}^{a\,\mu}_{\rho\, \pi}(\pi)$ and 
$\Delta \hat{J}^{a\,\mu}_{\rho\, \pi}(np)$
is
\be
\hat{J}^{a\,\mu}_{\rho\, \pi}(\pi) + \Delta \hat{J}^{a\,\mu}_{\rho\, \pi}(np)\,=\,
if_\pi q^\mu \Delta^\pi_F(q^2)\,\left[\hat{\cal M}^a_{\rho\, \pi}+\hat{\tilde {\cal M}}^a_{\rho\, \pi}\right]\,,
        \label{pc1}
\ee
whereas the other contact terms satisfy
\be
q_\mu \left[\hat{J}^{a\,\mu}_{c_1\,\pi}(a_1)+
\hat{J}^{a\,\mu}_{\rho\,\pi}(a_1)+\hat{J}^{a\,\mu}_{\rho\, \pi}\right]\,=\,if_\pi
\left[\hat{\cal M}^a_{\rho\, \pi}+\hat{\tilde {\cal M}}^a_{\rho\, \pi}\right]\,,
        \label{pc2}
\ee
as it should be, if the continuity equation (\ref{divnp1}) should hold also for the off-shell nucleons.

Evidently, the non-potential axial $\pi$ MEC operator is defined by Eq.\,(\ref{divnp1}) 
as
\be
\hat{J}^{a\,\mu}_{BS\,\pi}(np)\,=\,\hat{J}^{a\,\mu}_{c_1\,\pi}(a_1)+\hat{J}^{a\,\mu}_{\rho\,\pi}(a_1)+
\hat{J}^{a\,\mu}_{\rho\, \pi}(\pi)+\hat{J}^{a\,\mu}_{\rho\, \pi}+
\Delta\hat{J}^{a\,\mu}_{\rho\,\pi}(np)\,.  \label{NP1PIMEC}
\ee
As noted above, the natural place for the current $\hat{J}^{a\,\mu}_{c_1\,\pi}(a_1)$ is among the
non-potential ones.

We now consider another contribution to the non-potential currents of the pion range coming
from the $\Delta(1236)$ excitation currents. These currents appear because of vertices 
$a_1\,N \Delta$ and $\pi N \Delta$ and they are of the form
\bea
\hat{J}^{a\,\mu}_\pi(\Delta,\,a_1)\,&=&\,i f_\pi \left( \frac{f_{\pi N \Delta}}{m_\pi} \right) ^2\, m^2_{a_1}\,
\Delta^{\,\mu}_{a_1,\,\nu}(q)\, \hat{O}^{a\,\nu}_\pi(\Delta)\,,   \label{JDpia1} \\
\hat{J}^{a\,\mu}_\pi(\Delta,\,\pi)\,&=&\,i f_\pi q^\mu\, \Delta^\pi_F(q)\, \hat{\cal M}^a_\pi(\Delta)\,, \label{JDpipi} \\
\hat{\cal M}^a_\pi(\Delta)\,&=&\, \left( \frac{f_{\pi N \Delta}}{m_\pi} \right) ^2\,q_\nu\,
\hat{O}^{a\,\nu}_\pi(\Delta)\,, \label{Mpi} \\
\hat{O}^{a\,\nu}_\pi(\Delta)\,&=&\,\left[\,(T^+)^n T^a S^{\lambda \nu}_F(P)\,+\,
(T^+)^a T^b S^{\nu \lambda}_F(Q)\,\right]_1\,q_{2\,\lambda}\,\Delta^\pi_F(q_2)\,
\hat{\Gamma}^n_2\,+\,\ot\,. \label{O}
\eea
Here the propagator of the $\Delta$ isobar is defined as
\be
S^{\alpha \beta}_F(p)\,=\,\frac{1}{\not p - M_\Delta + i\varepsilon}\,\left[g^{\alpha \beta} - 
\frac{1}{3}\gamma^\alpha \gamma^\beta - \frac{2}{3 M^2_\Delta} p^\alpha p^\beta 
+ \frac{1}{3 M_\Delta} (p^\alpha \gamma^\beta - p^\beta \gamma^\alpha)\right]\,,  \label{Dp}
\ee
and the transition isospin operator $\vec T$ is defined in accord with Eq.\,(\ref{TS}). The total
$\Delta$ isobar current of the pion range is
\be
\hat{J}^{a\,\mu}_\pi(\Delta)\,=\,\hat{J}^{a\,\mu}_\pi(\Delta,\,a_1)\,
+\,\hat{J}^{a\,\mu}_\pi(\Delta,\,\pi)\,,\label{JDpi}
\ee
with the divergence
\be
q_\mu\,\hat{J}^{a\,\mu}_\pi(\Delta)\,=\,i f_\pi \,m^2_\pi\,\Delta^\pi_F(q)\,\hat{\cal M}^a_\pi(\Delta)\,.
\label{divJDpi}
\ee

We define the BS weak axial $\pi$ MEC operator as
\be
\hat{J}^{a\,\mu}_{BS\,\pi}(ex)\,=\,\hat{J}^{a\,\mu}_{BS\,\pi}(p)\,+\,\hat{J}^{a\,\mu}_{BS\,\pi}(np)
\,+\,\hat{J}^{a\,\mu}_\pi(\Delta)\,, \label{JPEXBS}
\ee
with $\hat{J}^{a\,\mu}_{BS\,\pi}(p)$, $\hat{J}^{a\,\mu}_{BS\,\pi}(np)$ 
and $\hat{J}^{a\,\mu}_\pi(\Delta)$ given in Eqs.\,(\ref{PPIMEC}), 
(\ref{NP1PIMEC}) and (\ref{JDpi}), respectively. Using Eqs.(\ref{divp2}), (\ref{divnp1})
and (\ref{divJDpi}) we obtain the WTI for the $\pi$ MEC operator
\be
q_\mu \hat{J}^{a\,\mu}_{BS\,\pi}(ex)\,=\,\left[\,\hat{e}_A(1)+\hat{e}_A(2),\,\hat{V}_\pi\,\right]_+\,
+\,if_\pi\,m^2_\pi\Delta^\pi_F(q^2)\,\hat{\cal M}^a_\pi(2)\,,  \label{divex}
\ee
where the pion absorption amplitude $\hat{\cal M}^a_\pi(2)$ is
\be
\hat{\cal M}^a_\pi(2)\,=\,\hat{\cal M}^a_{c_1\,\pi}
+2\Delta \hat{\cal M}^a_{c_1\,\pi}+\hat{\cal M}^a_{c_2\,\pi}+
\hat{\cal M}^a_{\rho\, \pi}+\hat{\tilde {\cal M}}^a_{\rho\, \pi}+\hat{\cal M}^a_\pi(\Delta)\,.  \label{Mapi2}
\ee

At low energies, the weak axial potential $\pi$ MECs play little role. The non-potential
current (\ref{Jrp}) dominates the time component of the weak axial $\pi$ MEC
and its effect can be clearly seen in light nuclei \cite{KDR},\cite{KT}-\cite{Ri}. On the other side, 
the current (\ref{JDpia1}) contributes significantly to the space component of this current,
as observed in the weak reactions in lightest nuclei, particularly in the reaction (\ref{mu3He})
\cite{Vo,CT}.

At higher energies, as in the case of the electromagnetic interaction, heavier mesons are expected 
to play a non-negligible role. Now we consider the weak axial MECs
of the $\rho$, $a_1$ and $\omega$ range.

\subsection{The axial MEC operator of the $\rho$ and $a_1$ meson range \label{CH22}}

As it will become clear soon the $\rho$ and $a_1$ exchanges should be considered 
in the given model simultaneously.
The weak axial MECs of the $\rho$ meson range derived from our model Lagrangian ${\cal L}^{YM}$ are as follows.
The only contact term is
\be
\hat{J}^{a\,\mu}_{c\,\rho}(\pi)\,=\,i f_\pi q^\mu\,\Delta^\pi_F(q^2)\,
\hat{\cal M}^a_{c\,\rho}\,,  \label{Jacrp}
\ee
where
\be
\hat{\cal M}^a_{c\,\rho}\,=\,\frac{g_A}{f_\pi} \left( \frac{g_\rho}{2} \right) ^2\,\frac{\kappa^V_\rho}{2M}
\delta_{a\,n}(\g5 \sigma_\eta^{\,\,\delta})_1\,q_{2 \delta}\Delta^{\eta\,\lambda}_\rho(q_2)
\hat{\Gamma}^n_{2\,\lambda}(q_2)\,+\,\ot\,. \label{Macrp}
\ee
The mesonic currents are
\bea
\hat{J}^{a\,\mu}_{a_1\,\rho}(a_1)\,&=&\,i\frac{g_A}{2} g_\rho^2 m_\rho^2 \,
\left[\,(q_\nu+q_{1\,\nu})\Delta^{\,\mu\zeta}_{a_1}(q)\,\Delta^{\,\,\,\,\lambda}_{a_1,\,\zeta}(q_1) 
-q_\zeta \Delta^{\,\,\,\,\mu}_{a_1,\,\nu}(q) \Delta^{\,\zeta \lambda}_{a_1}(q_1) \right. \nonumber  \\
& & \left. -q_{1 \zeta}\Delta^{\,\mu \zeta}_{a_1}(q) \Delta^{\,\,\,\,\lambda}_{a_1,\,\nu}(q_1)\,\right]
\,\vamn\,\hat \Gamma^{5\,m}_{1\,\lambda}\,\Delta^{\,\nu \eta}_{\rho}(q_2)\,\hat{\Gamma}^n_{2\,\eta}(q_2)\,
+\,\ot\,,   \label{Jaa1ra1}
\eea
and
\be
\hat{J}^{a\,\mu}_{a_1\,\rho}(\pi)\,=\,i f_\pi q^\mu\,\Delta^\pi_F(q^2)\,
\hat{\cal M}^a_{a_1\,\rho}(\pi)\,,  \label{Jaa1rp}
\ee
with
\bea
\hat{\cal M}^a_{a_1\,\rho}(\pi)\,&=&\,-\frac{g_A}{f_\pi} \left( \frac{g_\rho}{2} \right) ^2\, 
\left[\,(2q_{2\,\zeta}+q_{1\,\zeta})\,\Delta^{\,\,\,\,\lambda}_{a_1,\,\nu}(q_1)\,-\,
(2q_{2\,\nu}+q_{1\,\nu})\,\Delta^{\,\,\,\,\lambda}_{a_1,\,\zeta}(q_1)\,\right] \times \nonumber \\
& &\vamn\, \hat \Gamma^{5\,m}_{1\,\lambda}\,q^{\,\,\zeta}_2 \Delta^{\,\nu \eta}_{\rho}(q_2)\,
\hat{\Gamma}^n_{2\,\eta}(q_2)\,+\,\ot\,.   \label{Maa1rp}
\eea
Here we modify the notation for the pion absorption amplitude from $\hat{\cal M}^a_{a_1\,\rho}$ to
$\hat{\cal M}^a_{a_1\,\rho}(\pi)$.

The operator for the $\rho$ meson exchange potential is
\be
\hat{V}_\rho\,=\,\left (\frac{g_\rho}{2}\right )^2 \, \hat{\Gamma}^n_{1\,\mu}(-q_2)\,
\Delta^{\,\,\mu \nu}_\rho(q_2)\,\hat{\Gamma}^n_{2\,\nu}(q_2)\,.  \label{Vr}  
\ee
For the current of Eq.\,(\ref{Jacrp}), we can write the divergence in the form
\be
q_\mu \hat{J}^{a\,\mu}_{c\,\rho}(\pi)\,=\,i f_\pi m^2_\pi \Delta^\pi_F(q^2)
\,\hat{\cal M}^a_{c\,\rho}\,-\,i f_\pi \,\hat{\cal M}^a_{c\,\rho}\,. \label{dJacrp}
\ee
Using the explicit form of the amplitude $\hat{\cal M}^a_{c\,\rho}$ from Eq.\,(\ref{Macrp}),
the second term in Eq.\,(\ref{dJacrp}) can be cast into the form
\be
-\,i f_\pi \,\hat{\cal M}^a_{c\,\rho}\,=\,  \left[\hat{e}_A(1)\,+\,\hat{e}_A(2)\,,\,
\hat{V}^{an}_\rho \right]_+\,,  \label{acran}
\ee
where $\hat{V}^{an}_\rho$ is the part of the potential $\hat{V}_\rho$ given by the anomalous part
of the $\rho$NN vertex entering the anticommutator of the potential with the nucleon axial charge.
We denote the missing part of the anticommutator by $-\hat{\Delta}_\rho$. It is of the form
\be
-\hat{\Delta}_\rho\,\equiv\,\left[\hat{e}_A(1)\,+\,\hat{e}_A(2)\,,\,\hat{V}^n_\rho \right]_+\,=\,
ig_A\left(\frac{g_\rho}{2}\right)^2 \vamn\,\hat \Gamma^{5\,m}_{1\,\mu}\,
\Delta^{\,\,\mu \nu}_\rho(q_2)\,\hat{\Gamma}^n_{2\,\nu}(q_2)\,+\,\ot\,.  \label{mDr}  
\ee
In the next step of deriving the WTI for the $\rho$ meson MECs we calculate
\bea
q_\mu \hat{J}^{a\,\mu}_{a_1\,\rho}(a_1)+\hat{\Delta}_\rho&=&
ig_A\left(\frac{g_\rho}{2}\right)^2 \vamn \hat \Gamma^{5\,m}_{1\,\lambda}\,
\Delta^{a_1}_F(q^2_1)\,\left [ q_{1\,\nu}q^\lambda - (q_1 \cdot q_2 + m^2_{a_1})
g^{\,\,\lambda}_\nu \right ]  \nonumber  \\
& & \times\,\Delta^{\,\,\nu \eta}_\rho(q_2)\,\hat{\Gamma}^n_{2\,\eta}(q_2)   
\,+\,\ot\,\equiv\,\,i f_\pi \,\hat{\cal M}^a_{a_1\,\rho}(x)\,. \label{dJa1pdr}  
\eea
However, the amplitude $\hat{\cal M}^a_{a_1\,\rho}(x)$ does not coincide with the amplitude
$\hat{\cal M}^a_{a_1\,\rho}(\pi)$ from Eq.\,(\ref{Maa1rp}) and  we get
\bea
q_\mu \left(\hat{J}^{a\,\mu}_{a_1\,\rho}(a_1)\,+\,\hat{J}^{a\,\mu}_{a_1\,\rho}(\pi)\right)\,&=&\,
i f_\pi m^2_\pi\,\Delta^\pi_F(q^2)\,\hat{\cal M}^a_{a_1\,\rho}(\pi)\,+\,
 \left[\hat{e}_A(1)\,+\,\hat{e}_A(2)\,,\,\hat{V}^n_\rho \right]_+\,  \nonumber  \\
& & + i f_\pi \Delta^\pi_F(q^2)\,\left(\hat{\cal M}^a_{a_1\,\rho}(x)\,-\,
\hat{\cal M}^a_{a_1\,\rho}(\pi)\right)\,,  \label{dJaa1r}
\eea
which together with Eqs.\,(\ref{dJacrp}) and (\ref{acran}) provides for the
exchange current $\hat{\tilde J}^{a\,\mu}_{BS\,\rho}(ex)$, defined as
\be
\hat{\tilde J}^{a\,\mu}_{BS\,\rho}(ex)\,=\,\hat{J}^{a\,\mu}_{c\,\rho}(\pi)\,+\,
\hat{J}^{a\,\mu}_{a_1\,\rho}(a_1)\,+\,\hat{J}^{a\,\mu}_{a_1\,\rho}(\pi)\,,  \label{JTREXBS}
\ee
with the WTI in the form
\be
q_\mu \hat{\tilde J}^{a\,\mu}_{BS\,\rho}(ex)= \left[\hat{e}_A(1)+\hat{e}_A(2),\hat{V}_\rho \right]_+
+i f_\pi m^2_\pi\,\Delta^\pi_F(q^2)\hat{ {\tilde {\cal M}}}^a_\rho(2)
+ i f_\pi\left(\hat{\cal M}^a_{a_1\,\rho}(x)-
\hat{\cal M}^a_{a_1\,\rho}(\pi)\right)\,,
  \label{dJTREXBS}
\ee
where the pion absorption amplitude is given as
\be
\hat{\tilde{ {\cal M}}}^a_\rho(2)\,=\,\hat{\cal M}^a_{c\,\rho}\,+\,\hat{\cal M}^a_{a_1\,\rho}(\pi)\,. 
\label{Mtar2}
\ee
Using Eqs.\,(\ref{Maa1rp}) and (\ref{dJa1pdr}), we derive for 
the difference of the amplitudes
$\hat{\cal M}^a_{a_1\,\rho}(x)$ and $\hat{\cal M}^a_{a_1\,\rho}(\pi)$ the following equation
\be
if_\pi\,\left(\hat{\cal M}^a_{a_1\,\rho}(x)\,-\,\hat{\cal M}^a_{a_1\,\rho}(\pi)\right)\,=
\,-ig_A\,\frac{g^2_\rho}{2}\,\vamn\,
\hat{\Gamma}_{1\,m}^{\,5\,\mu}\,\Delta_{a_1}^{\,\mu \nu}(q_1)\,
\hat{\Gamma}^n_{2\,\nu}(q_2)\,+\,\ot\,.   \label{difMxMpi}
\ee
Evidently, this difference is an operator of the $a_1$ meson range.
As we shall see below, it will be compensated by the contribution from the weak axial
exchange currents of the $a_1$ meson range.

Considered together with the $\rho$ meson MECs, the $a_1$ meson MECs contain only 
3 contact terms $\hat {J}^{a\,\mu}_{c_i\,a_1}(\pi)$ (i=1,2,3)
\be
\hat {J}^{a\,\mu}_{c_i\,a_1}(\pi)\,=\,if_\pi\,q^\mu\,\Delta^\pi_F(q^2)\,
\hat{\cal M}^a_{c_i\,a_1}\,. \label{Jacia1}
\ee
The associated pion absorption amplitudes are
\bea
\hat{\cal M}^a_{c_1\,a_1}\,&=&\,-\frac{g_A}{f_\pi}\left(g_A g_\rho\right)^2\,\vamn
(\gamma_\mu \tau^m)_1\,\Delta^{\,\mu \nu}_{a_1}(q_2)\,
\hat{\Gamma}^{\,5\,n}_{2\,\nu}\,+\,\ot\,,  \label{Mac1a1} \\
\hat{\cal M}^a_{c_2\,a_1}\,&=&\,\frac{g_A}{f_\pi}\frac{g^2_\rho}{2}\,\vamn\,
(\gamma_\mu \tau^m)_1\,\Delta^{\,\mu \nu}_{a_1}(q_2)\,
\hat{\Gamma}^{\,5\,n}_{2\,\nu}\,+\,\ot\,,   \label{Mac2a1} \\
\hat{\cal M}^a_{c_3\,a_1}\,&=&\,-i\frac{g_A}{f_\pi}\,\frac{g^2_\rho}{2}\,\frac{\kappa^V_\rho}{2M}\,
\vamn\,q_{1\,\mu}\left(\sigma^\mu_{\,\nu}\tau^m\right)_1\,\Delta^{\nu \lambda}_{a_1}(q_2)\,
\hat{\Gamma}^{\,5\,n}_{2\,\lambda}\,+\,\ot\,.    \label{Mac3a1}
\eea
Let us note that the currents (\ref{Jacia1}) are derived from the same NN$\pi a_1$ vertices 
as the $\pi$ MECs (\ref{Jnpa})-(\ref{JPCAC}), only the role of the weakly and strongly
interacting mesons is exchanged. 

The $a_1$ meson exchange potential is
\be
\hat V_{a_1}\,=\,\left(g_A g_\rho\right)^2\,
\hat{\Gamma}^{\,5\,n}_{1\,\mu}\,\Delta^{\,\mu \nu}_{a_1}(q_2)\,\hat{\Gamma}^{\,5\,n}_{2\,\nu}\,.
\label{Va1}
\ee

Let us now calculate the divergence of the currents $\hat {J}^{a\,\mu}_{c_i\,a_1}(\pi)$.
In the first step we compute
\be
q_\mu\hat {J}^{a\,\mu}_{c_1\,a_1}(\pi)\,=\,if_\pi\,m^2_\pi\,\Delta^\pi_F(q^2)\,
\hat{\cal M}^a_{c_1\,a_1}\,-\,if_\pi\,\hat{\cal M}^a_{c_1\,a_1}\,.   \label{dJac1a1}
\ee
Using the explicit form (\ref{Mac1a1}) of $\hat{\cal M}^a_{c_1\,a_1}$ it follows that
\be
-\,if_\pi\,\hat{\cal M}^a_{c_1\,a_1}\,=\,\left[\hat{e}_A(1)\,+\,\hat{e}_A(2)\,,\,\hat{V}_{a_1}\right]_+\,.  \label{ceAVa1}
\ee

We next calculate
\be
q_\mu\left(\hat {J}^{a\,\mu}_{c_2\,a_1}(\pi)+\hat {J}^{a\,\mu}_{c_3\,a_1}(\pi)\right)=
if_\pi\,m^2_\pi\,\Delta^\pi_F(q^2)\left(\hat{\cal M}^a_{c_2\,a_1} + \hat{\cal M}^a_{c_3\,a_1}\right)
-if_\pi \left(\hat{\cal M}^a_{c_2\,a_1} + \hat{\cal M}^a_{c_3\,a_1}\right)\,.   \label{dJac2pc3a1}
\ee
Employing Eqs.\,(\ref{dJac1a1})--(\ref{dJac2pc3a1}), we have the following WTI 
for the weak axial $a_1$ MEC derived so far 
\be
q_\mu \hat{\tilde J}^{a\,\mu}_{BS\,a_1}(ex)=\left[\hat{e}_A(1)\,+\,\hat{e}_A(2)\,,\,\hat{V}_{a_1} \right]_+
+ i f_\pi m^2_\pi\,\Delta^\pi_F(q^2)\hat{ {\tilde {\cal M}}}^a_{a_1}(2)
- i f_\pi\left(\hat{\cal M}^a_{c_2\,a_1}+
\hat{\cal M}^a_{c_3\,a_1}\right)\,,  \label{dJTA1EXBS}
\ee
with the current $\hat{\tilde J}^{a\,\mu}_{BS\,a_1}(ex)$ defined as
\be
\hat{\tilde J}^{a\,\mu}_{BS\,a_1}(ex)\,=\,\sum\limits^3_{i=1}\,\hat {J}^{a\,\mu}_{c_i\,a_1}(\pi)\,,
\label{JRA1EXBS}
\ee
and with the pion absorption amplitude $\hat{ {\tilde {\cal M}}}^a_{a_1}(2)$ given by
\be
\hat{\tilde{ {\cal M}}}^a_{a_1}(2)\,=\,\sum\limits^3_{i=1}\hat{\cal M}^a_{c_i\,a_1}\,.  \label{Mtaa12}
\ee
The sum of the last two terms at the right hand side of Eq.\,(\ref{dJTA1EXBS}) yields
\be
-\,if_\pi\,\left(\hat{\cal M}^a_{c_2\,a_1}\,+\,\hat{\cal M}^a_{c_3\,a_1}\right)\,=\,
ig_A\,\frac{g^2_\rho}{2}\,\vamn\,
\hat{\Gamma}_{1\,m}^{\,5\,\mu}\,\Delta_{a_1}^{\,\mu \nu}(q_1)\,
\hat{\Gamma}^n_{2\,\nu}(q_2)\,+\,\ot\,.   \label{difMc2Mc3}
\ee
Observing that the right hand sides of Eqs.\,(\ref{difMxMpi}) and (\ref{difMc2Mc3}) are of the same 
form but with the opposite sign we derive  the WTI for the sum of the 
weak axial $\rho$ and $a_1$ MECs 
\bea
q_\mu\left(\hat{\tilde J}^{a\,\mu}_{BS\,\rho}(ex)\,+\,\hat{\tilde J}^{a\,\mu}_{BS\,a_1}(ex)\,\right)\,&=&\,
 \left[\hat{e}_A(1)\,+\,\hat{e}_A(2)\,,\,\hat{V}_\rho\,+\,\hat{V}_{a_1} \right]_+\,  \nonumber \\
&  & \,+i f_\pi m^2_\pi\,\Delta^\pi_F(q^2)\,\left(\hat{ {\tilde {\cal M}}}^a_\rho(2)\,
+\,\hat{ {\tilde {\cal M}}}^a_{a_1}(2)\right)\,.   \label{dJTRPA1EXBS}
\eea
Using  Eqs.\,(\ref{difMxMpi}) and (\ref{difMc2Mc3}) once more, we get from Eqs.\,(\ref{Mtar2})
and (\ref{Mtaa12})
\be
\hat{ {\tilde {\cal M}}}^a_\rho(2)\,+\,\hat{ {\tilde {\cal M}}}^a_{a_1}(2)\,=\,
\hat{\cal M}^a_{c\,\rho}\,+\,\hat{\cal M}^a_{a_1\,\rho}(x)\,+\,
\hat{\cal M}^a_{c_1\,a_1}\,.  \label{Mara12}
\ee

We now derive the $\Delta$ excitation currents of the $\rho$ and $a_1$ range.
In analogy with Eqs.(\ref{JDpia1})--(\ref{divJDpi}) we have for the $\rho$ MEC the following set of equations
\bea
\hat{J}^{a\,\mu}_\rho(\Delta,\,a_1)\,&=&\,-f_\pi \frac{f_{\pi N \Delta}}{m_\pi}\,
\frac{G_1}{M}\,\left(g_\rho m_\rho\right)^2\,
\Delta^{\,\mu}_{a_1,\,\nu}(q)\, \hat{O}^{a\,\nu}_\rho(\Delta)\,,   \label{JDrhoa1} \\
\hat{J}^{a\,\mu}_\rho(\Delta,\,\pi)\,&=&\,i f_\pi q^\mu\, \Delta^\pi_F(q)\, 
\hat{\cal M}^a_\rho(\Delta)\,, \label{JDrhopi} \\
\hat{\cal M}^a_\rho(\Delta)\,&=&\, i\frac{f_{\pi N \Delta}}{m_\pi}\,\frac{G_1}{M}\,
\frac{g^2_\rho}{2}\,q_\nu\,\hat{O}^{a\,\nu}_\rho(\Delta)\,, \label{Mrho} \\
\hat{O}^{a\,\nu}_\rho(\Delta)\,&=&\,\left[\,(T^+)^n T^a \g5 \gamma_\zeta S^{\lambda \nu}_F(P)\,+\,
(T^+)^a T^n S^{\nu \lambda}_F(Q)\g5 \gamma_\zeta\,\right]_1\,
\left(q_{2\,\lambda}\,g^{\zeta \eta}\, \right. \nonumber \\
& &\left. -\,q_2^{\,\zeta}\,g_\lambda^{\,\,\eta}\right)\,\Delta^\rho_F(q_2)\,
\hat{\Gamma}^n_{2\,\eta}(q_2)\,+\,\ot\,. \label{Orho}
\eea
The total $\Delta$ isobar current of the $\rho$ meson range is
\be
\hat{J}^{a\,\mu}_\rho(\Delta)\,=\,\hat{J}^{a\,\mu}_\rho(\Delta,\,a_1)\,
+\,\hat{J}^{a\,\mu}_\rho(\Delta,\,\pi)\,,\label{JDrho}
\ee
with the divergence
\be
q_\mu\,\hat{J}^{a\,\mu}_\rho(\Delta)\,=\,i f_\pi \,m^2_\pi\,\Delta^\pi_F(q^2)\,\hat{\cal M}^a_\rho(\Delta)\,.
\label{divJDrho}
\ee
Similarly we have for the $\Delta$ excitation currents of the $a_1$ meson range
\bea
\hat{J}^{a\,\mu}_{a_1}(\Delta,\,a_1)\,&=&\,g_A\left(2f_\pi g_\rho m_\rho\right)^2\,
\left(\frac{f_{\pi N \Delta}}{m_\pi}\right)^2\,
\Delta^{\,\mu}_{a_1,\,\nu}(q)\, \hat{O}^{a\,\nu}_{a_1}(\Delta)\,,   \label{JDa1a1} \\
\hat{J}^{a\,\mu}_{a_1}(\Delta,\,\pi)\,&=&\,i f_\pi q^\mu\, \Delta^\pi_F(q)\, 
\hat{\cal M}^a_{a_1}(\Delta)\,, \label{JDa1pi} \\
\hat{\cal M}^a_{a_1}(\Delta)\,&=&\, -2i g_A f_\pi g^2_\rho\,
\left(\frac{f_{\pi N \Delta}}{m_\pi}\right)^2\,
q_\nu\,\hat{O}^{a\,\nu}_{a_1}(\Delta)\,, \label{Ma1} \\
\hat{O}^{a\,\nu}_{a_1}(\Delta)\,&=&\,\left[\,(T^+)^n T^a  S^{\lambda \nu}_F(P)\,+\,
(T^+)^a T^n S^{\nu \lambda}_F(Q)\,\right]\,\Delta^{\,\,\eta}_{a_1,\,\,\lambda}(q_2)\,
\hat{\Gamma}^{5\,n}_{2\,\eta}\,+\,\ot\,. \label{Oa1}
\eea
The total $\Delta$ isobar current of the $a_1$ meson range is
\be
\hat{J}^{a\,\mu}_{a_1}(\Delta)\,=\,\hat{J}^{a\,\mu}_{a_1}(\Delta,\,a_1)\,
+\,\hat{J}^{a\,\mu}_{a_1}(\Delta,\,\pi)\,,\label{JDa1}
\ee
with the divergence
\be
q_\mu\,\hat{J}^{a\,\mu}_{a_1}(\Delta)\,=\,i f_\pi \,m^2_\pi\,\Delta^\pi_F(q^2)\,\hat{\cal M}^a_{a_1}(\Delta)\,.
\label{divJDa1}
\ee
The total BS weak axial $\rho$+$a_1$ MEC is defined as
\be
\hat{J}^{a\,\mu}_{BS\,\rho + a_1}(ex)\,=\,\hat{\tilde J}^{a\,\mu}_{BS\,\rho}(ex)\,
+\,\hat{\tilde J}^{a\,\mu}_{BS\,a_1}(ex)\,+\,\hat{J}^{a\,\mu}_\rho(\Delta)
\,+\,\hat{J}^{a\,\mu}_{a_1}(\Delta)\,, \label{JRPA1EXBS}
\ee
and it satisfies the WTI
\be
q_\mu \hat{J}^{a\,\mu}_{BS\,\rho + a_1}(ex)\,=\,
 \left[\hat{e}_A(1)\,+\,\hat{e}_A(2)\,,\,\hat{V}_\rho\,+\,\hat{V}_{a_1} \right]_+\, 
+\,i f_\pi m^2_\pi\,\Delta^\pi_F(q^2)\,\hat {\cal M}^a_{\rho + a_1}(2)\,,   \label{dJRPA1EXBS}
\ee
with the pion absorption amplitude $\hat{\cal M}^a_{\rho + a_1}(2)$ given by
\be
\hat{\cal M}^a_{\rho + a_1}(2)\,=\,\hat{\cal M}^a_{c\,\rho}\,+\,\hat{\cal M}^a_{a_1\,\rho}(x)\,+\,
\hat{\cal M}^a_{c_1\,a_1}\,+\,\hat{\cal M}^a_\rho(\Delta)\,+\,\hat{\cal M}^a_{a_1}(\Delta)\,. \label{Marpa1}
\ee

The last MEC operator we will now construct is that of the $\omega$ meson range. 

\subsection{The axial MEC operator of the $\omega$ meson range \label{CH23}}

The $\omega$ meson is a vector meson with the isospin zero and the only MEC operator
which appears is the contact current $\hat J^{a\,\mu}_{c\,\omega}(\pi)$
\bea
\hat J^{a\,\mu}_{c\,\omega}(\pi)\,&=&\,if_\pi\,q_\mu\,\Delta^\pi_F(q)\,\hat{\cal M}^a_{c\,\omega}\,, \label{Jacop} \\
\hat{\cal M}^a_{c\,\omega}\,&=&\,\frac{g_a}{f_\pi}\left(\frac{g_\omega}{2}\right)^2\,
\frac{\kappa^S}{2M}\,\left(\g5\,\sigma^{\,\,\nu}_{\mu}\,q_{2\,\nu}\,\tau^a\right)_1\,
\Delta^{\,\mu \eta}_\omega(q_2)\,\hat \Gamma_{2\,\eta}(q_2)\,+\,\ot\,.   \label{Macop}
\eea
Here $\hat \Gamma_{2\,\eta}(q_2)$ is given by Eq.\,(\ref{Gmin}) with the isospin operator $\tau^m$ omitted
and with the change $\kappa^V_\rho\,\rightarrow\,\kappa^S$.

The $\omega$ exchange potential reads
\be
\hat V_\omega\,=\,\left (\frac{g_\omega}{2}\right )^2 \, \hat{\Gamma}_{1\,\mu}(-q_2)\,
\Delta^{\,\,\mu \nu}_\omega(q_2)\,\hat{\Gamma}_{2\,\nu}(q_2)\,.  \label{Vo}  
\ee
Calculating the divergence of the current (\ref{Jacop}) yields
\be
q_\mu \hat J^{a\,\mu}_{c\,\omega}(\pi)\,=\,if_\pi\,m^2\pi\,\Delta^\pi_F(q^2)\,
\hat{\cal M}^a_{c\,\omega}\,-\,if_\pi\,\hat{\cal M}^a_{c\,\omega}\,.  \label{dpJacop}
\ee
Using  Eq.\,(\ref{Macop}) for the amplitude $\hat{\cal M}^a_{c\,\omega}$ and Eq.\,(\ref{Vo})
for the $\omega$ exchange potential $\hat V_\omega$ we verify that the second term on 
the right hand side of Eq.\,(\ref{dpJacop}) provides the needed anticommutator of the
nucleon axial charge and the potential. Finally we have the WTI for the BS weak axial $\omega$ MEC
\be
q_\mu \hat J^{a\,\mu}_{c\,\omega}(\pi)\,=\,if_\pi\,m^2\pi\,\Delta^\pi_F(q^2)\,
\hat{\cal M}^a_{c\,\omega}\,+\,\left[\hat{e}_A(1)\,+\,\hat{e}_A(2)\,,\,\hat{V}_\omega\right]_+\,.  \label{dJacop}
\ee

We have completed the derivation of the BS weak axial one-body operator and of the $\pi$, $\rho$, $a_1$ 
and $\omega$ MEC operators in the tree approximation.
These operators satisfy separately the WTI (\ref{DOBCi}),(\ref{divex}),
(\ref{dJRPA1EXBS}) and (\ref{dJacop}). Starting from them,
we verify in the next step that the matrix element of the total current satisfies the standard PCAC equation .

\subsection{The continuity equation for the current matrix element \label{CH24}}

We define, in accord with Sect.\,2 of Ref.\,\cite{AVOG} the full BS weak axial current as 
\be
\hat{J}^{a\,\mu}_{BS}\,=\,i\hat{J}^{\mu}_a(1,1)G^{-1}_2\,+\,i\hat{J}^{\mu}_a(1,2)G^{-1}_1
\,+\,\hat{J}^{a\,\mu}_{BS}(ex)\,=\,\hat{J}^{a\,\mu}_{IA}\,+\,\hat{J}^{a\,\mu}_{BS}(ex)\,, \label{JBSE}
\ee
where 
\be
\hat{J}^{a\,\mu}_{BS}(ex)\,=\,\hat{J}^{a\,\mu}_{BS\,\pi}(ex)\,+\,\hat{J}^{a\,\mu}_{BS\,\rho+a_1}(ex)\,
+\,\hat{J}^{a\,\mu}_{c\,\omega}(\pi)\,.   \label{JBSEX}
\ee
Using the WTIs  for the one- and two-nucleon currents yields for the
divergence of the full BS current
\be
q_\mu \hat{J}^{a\,\mu}_{BS}\,=\,[\,\hat{e}_A(1)+\hat{e}_A(2),\,{\cal G}^{-1}\,]_+\,
+\,if_\pi\,m^2_\pi\Delta^\pi_F(q^2)\,\hat{\cal M}^a\,,  \label{divf}
\ee
where the inverse Green function is 
\be
{\cal G}^{-1}\,=\,G^{-1}_{BS}\,+\,\hat V\,,   \label{IGF}
\ee
with
\be
\hat V\,=\,\hat V_\pi\,+\,\hat V_\rho\,+\,\hat V_{a_1}\,+\,\hat V_\omega\,.  \label{V}
\ee
The BS propagator in term of the single-particle propagators reads $G_{BS}=-iG_1 G_2$ and
\be
\hat{\cal M}^a\,=\,i\hat{\Gamma}^a_1 G^{-1}_2+i\hat{\Gamma}^a_2 G^{-1}_1+
\hat{\cal M}^a_\pi(2)+\hat{\cal M}^a_{\rho+a_1}(2)+\hat{\cal M}^a_{c\,\omega}\,.   \label{Ma}
\ee
Because the two-body BS wave functions for both bound and scattering states satisfy
the equation
\be
{\cal G}^{-1}|\psi>\,=\,<\psi|{\cal G}^{-1}\,=\,0\,,  \label{BSE}
\ee
the matrix element of the divergence of the BS current (\ref{divf}) fulfil
the standard PCAC constraint
\be
q_\mu <\psi|\,\hat{J}^{a\,\mu}_{BS}\,|\psi>\,=\,if_\pi\,m^2_\pi\Delta^\pi_F(q^2)\,
<\psi|\,\hat{\cal M}^a\,|\psi>\,.  \label{divme}
\ee

\subsection{The strong form factors \label{CH25}}

We have dealt with the point BNN vertices so far. 
The usual way to introduce the strong interaction effects in the BNN vertices is to 
introduce the form factors F$_{BNN}$(q$^2_i$) with the normalization F$_{BNN}$(m$^2_B$)=1.
Then the vertex $\hat{\Gamma}^a_i$ defined in (\ref{vpvps}) becomes
\be
\hat{\Gamma}^a_i\,\longrightarrow\,\hat{\Gamma}^a_i\, F_{\pi NN}(q^2_i)\,. \label{piNNFF}
\ee
It is a simple matter to verify that the one-nucleon current (\ref{ONAC}) should be redefined
similarly
\be
\hat{J}^{a\,\mu}(1,i)\,\longrightarrow\,\hat{J}^{a\,\mu}_a(1,i)\, F_{\pi NN}(q^2_i)\,. \label{OBACNFF}
\ee
Then for the divergence of this current we have the same Eq.\,(\ref{DOBCi}) but with 
the vertex $\hat{\Gamma}^a_i$ from (\ref{piNNFF}) and with the nucleon axial charge 
$\hat{e}_A(i)$ redefined according to the prescription (\ref{piNNFF})
\be
\hat{e}_A(i)\,\longrightarrow\,\hat{e}_A(i)\,F_{\pi NN}(q^2_i)\,.
\ee
We could also leave the axial charge unchanged but then we should redefine the propagator in a similar manner
\be
G^{-1}(\,p\,)\,\longrightarrow\,G^{-1}(\,p\,)\,F_{\pi NN}(q^2_i)\,\equiv\,
\not p - M - \Sigma\,,  \label{GmONFF}
\ee
with
\be
\Sigma\,=\,(\not p - M)[1-F_{\pi NN}(q^2_i)]\,.
\ee

Introducing the strong form factors into the potential currents (\ref{JPCAC}) and 
(\ref{Jppp})-(\ref{Jpp}) and to the $\Delta$ excitation currents (\ref{JDpia1})--(\ref{JDpipi}) 
follows the standard prescription \cite{GR} and the pion propagator
connecting the baryon lines is modified as
\be
\Delta^\pi_F(q^2_i)\,\longrightarrow\,\Delta^\pi_F(q^2_i)\,
F^2_{\pi NN}(q^2_i)\,.  \label{Dpim}
\ee
This prescription is not suitable for the non-potential currents (\ref{Jpra})-(\ref{Jrp}).
In this case the correct procedure is
\be
\Delta^B_F(q^2_i)\,\longrightarrow\,\Delta^B_F(q^2_i)\,F_{BNN}(q^2_i)\,,
\quad B\,=\,\pi,\,\rho\,.  \label{DBm}
\ee
In order that the WTI (\ref{divnp1}) would be satisfied in the presence of the 
strong form factors, the pion propagator of the current (\ref{Jnpa}) should be modified as
\be
\Delta^\pi_F(q^2_i)\,\longrightarrow\,\Delta^\pi_F(q^2_i)\,F_{\pi NN}(q^2_i)
F_{\rho NN}(q^2_j)\,\,.  \label{Dpip}
\ee
Checking the WTI for operators of the $\rho$ and $a_1$ meson range we conclude that it
would be valid also with the strong form factors included provided 
\begin{enumerate}
\item In all  operators of the $\rho$ range
\be
\Delta^\rho_F(q^2_i)\,\longrightarrow\,\Delta^\rho_F(q^2_i)\,
F^2_{\rho NN}(q^2_i)\,.  \label{Drhom}
\ee
\item In the amplitudes $\hat{\cal M}^a_{c_2\,a_1}$ and $\hat{\cal M}^a_{c_3\,a_1}$
\be
\Delta^{a_1}_F(q^2_i)\,\longrightarrow\,\Delta^{a_1}_F(q^2_i)\,
F^2_{\rho NN}(q^2_i)\,.  \label{Da11m}
\ee
\item In the amplitude $\hat{\cal M}^a_{c_1\,a_1}$ and in the potential $\hat V_{a_1}$ and
in the currents $\hat{J}^{a\,\mu}_{a_1}(\Delta,\,a_1)$ and $\hat{J}^{a\,\mu}_{a_1}(\Delta,\,\pi)$
\be
\Delta^{a_1}_F(q^2_i)\,\longrightarrow\,\Delta^{a_1}_F(q^2_i)\,
F^2_{a_1 NN}(q^2_i)\,.  \label{Da12m}
\ee
\end{enumerate}
Finally it holds for the operators of the $\omega$ meson range that the necessary modification is
\be
\Delta^\omega_F(q^2_i)\,\longrightarrow\,\Delta^\omega_F(q^2_i)\,
F^2_{\omega NN}(q^2_i)\,.  \label{Dom}
\ee

In the next section, we investigate the model dependence of our MECs appearing due to
the difference in the physical content of the YM type minimal Lagrangian ${\cal L}^{M}$ and 
of the HLS Lagrangian ${\cal L}^{H}$.

\section{Model dependence of the BS MECs  \label{CH3}}

The derivation of the MECs from the HLS Lagrangians and currents (\ref{LhlsNPRA1O})-
(\ref{JAHLS}) is analogous to that of the previous sections. Therefore, we present
here briefly the results and emphasize the difference leading to the model dependence
of the currents. The currents of this section will be labeled by an additional label
H in order to distinguish them from the currents derived from the minimal Lagrangian
${\cal L}^M$.

\subsection{The axial MEC operator of the pion range \label{CH31}}

It is strightforward to check that the potential axial $\pi$ MEC operator $\hat{J}^{a\,\mu}_{BS\,\pi}(p,H)$
coincides with the current $\hat{J}^{a\,\mu}_{BS\,\pi}(p)$ of Eq.\,(\ref{PPIMEC}). 
But generally, the non-potential currents differ.  It holds only
for the new $\rho \pi \pi$ current $\hat{J}^{a\,\mu}_{\rho\,\pi}(\pi,H)$  that
\be
\hat{J}^{a\,\mu}_{\rho\,\pi}(\pi,H)\,=\,\hat{J}^{a\,\mu}_{\rho\,\pi}(\pi)\,, \label{Jrpph}
\ee
where $\hat{J}^{a\,\mu}_{\rho\,\pi}(\pi)$ is given in Eq.\,(\ref{Jrpp}). Checking further 
the currents $\vec{J}^{\,5\,\mu}_{M}$ of Eq.\,(\ref{OBYMCS}) and $\vec{J}^{\,5\,\mu}_{H}$
of Eq.\,(\ref{JAHLS}) we find that the new $\rho \pi$ contact term is now
\be
\hat{J}^{a\,\mu}_{\rho\,\pi}(1,H)\,=\,2\hat{J}^{a\,\mu}_{\rho\,\pi}\,,  \label{Jrp1h}
\ee
where the current $\hat{J}^{a\,\mu}_{\rho\,\pi}$ is given in Eq.\,(\ref{Jrp}). Here we label
the current (\ref{Jrp1h}) additionally, because another $\rho \pi$ term appears due to the
presence of the vertex $\sim$ $\partial^\nu \vec {\Pi} \times \vec {\rho}^{\,\mu}_{\,\,\nu}$ in the current 
$\vec{J}^{\,5\,\mu}_{H}$,
\be
\hat{J}^{a\,\mu}_{\rho\,\pi}(2,H)\,=\-\frac{1}{4f_\pi}\,\vamn\Gamma^m_{1\,\nu}(q_1)\left[ q^\mu_1\,q^\nu_2 - 
(q_1 \cdot q_2) g^{\mu \nu} \right]\, \Delta^\pi_F(q^2_2) \hat \Gamma^n_2\,+\,
\ot\,.  \label{Jrp2h}
\ee
The new current $\hat{J}^{a\,\mu}_{\rho\,\pi}(a_1,H)$, obtained using the Lagrangian
$\Delta{\cal L}^{H}_{\pi \rho a_{1}}$ of Eq.\,(\ref{DLHLSPRA1}) is
\bea
\hat{J}^{a\,\mu}_{\rho\,\pi}(a_1,H)\,&=&\,\frac{m^2_{a_1}}{4f_\pi}\,\Delta^{\,\,\mu}_{a_1\,\nu}(q)\,
\vamn \hat{\Gamma}^m_{1\,\eta}(q_1)\Delta^\rho_F(q^2_1)\left[q^\nu_1 q^\eta_2 - (q_1 \cdot q_2) g^{\nu\,\eta} 
\right]\,\Delta^\pi_F(q^2_2)\,\hat{\Gamma}^n_2  \nonumber  \\
& & \hspace{-50pt}+\,\frac{m^2_{a_1}}{8f_\pi}\,\Delta^{a_1}_F(q^2)\,
\vamn \hat{\Gamma}^m_{1\,\eta}(q_1)\Delta^{\,\eta}_{\rho\,\nu}(q_1)\,
\left[q^\nu q^\mu_2 - (q \cdot q_2) g^{\nu\,\mu} 
\right]\,\Delta^\pi_F(q^2_2)\hat{\Gamma}^n_2\,+\,\ot\,,  \label{Jprah}
\eea
The new non-potential currents satisfy the following WTIs
\be
q_\mu\,\left[ \hat{J}^{a\,\mu}_{\rho\,\pi}(a_1,H) + \hat{J}^{a\,\mu}_{\rho\,\pi}(2,H) \right]\,=\,0\,,
\label{wtih1}  
\ee
\be
q_\mu\,\left[ \hat{J}^{a\,\mu}_{\rho\,\pi}(1,H) + \hat{J}^{a\,\mu}_{\rho\,\pi}(\pi,H) +
\Delta\hat{J}^{a\,\mu}_{\rho\,\pi}(np) \right]\,=\,if_\pi\,m^2_\pi
\Delta^\pi_F(q^2)\,\left[\hat{\cal M}^a_{\rho\, \pi}+\hat{\tilde {\cal M}}^a_{\rho\, \pi}\right]\,,
        \label{wtih2}
\ee
where the current $\Delta\hat{J}^{a\,\mu}_{\rho\,\pi}(np)$ is given in Eq.\,(\ref{dJnp})
and the pion absorption amplitudes $\hat{\cal M}^a_{\rho\, \pi}$ and 
$\hat{\tilde {\cal M}}^a_{\rho\, \pi}$ are from Eqs.\,(\ref{Mrpp}) and (\ref{Mtrp}), respectively.
In analogy with the non-potential currents of the Sect.\, \ref{CH21}, in addition to Eq.\,(\ref{pc1})
we have now
\be
q_\mu\,\hat{J}^{a\,\mu}_{\rho\,\pi}(1,H)\,=\,if_\pi
\left[\hat{\cal M}^a_{\rho\, \pi}+\hat{\tilde {\cal M}}^a_{\rho\, \pi}\right]\,,
        \label{wtih3}
\ee
instead of Eq.\,(\ref{pc2}).

\subsection{The axial MEC operator of the $\rho$ and $a_1$ meson range \label{CH32}}

In this model, the $\rho$ and $a_1$ axial MECs satisfy the WTI separately.  We first consider the $\rho$ meson axial MECs.

Instead of the current $\hat{J}^{a\,\mu}_{c\,\rho}(\pi)$ of Eq.\,(\ref{Jacrp}),
we have now two contact terms $\hat{J}^{a\,\mu}_{c_1\,\rho}(\pi,H)$ and $\hat{J}^{a\,\mu}_{c_2\,\rho}(\pi,H)$
\be
\hat{J}^{a\,\mu}_{c_1\,\rho}(\pi,H)\,=\,\hat{J}^{a\,\mu}_{c\,\rho}(\pi)\,,  \label{Jac1rp}
\ee
and
\be
\hat{J}^{a\,\mu}_{c_2\,\rho}(\pi,H)\,=\,i f_\pi q^\mu\,\Delta^\pi_F(q^2)\,
\hat{\cal M}^a_{c_2\,\rho}(H)\,,  \label{Jac2rp}
\ee
where the pion absorption amplitude $\hat{\cal M}^a_{c_2\,\rho}(H)$ is
\be 
\hat{\cal M}^a_{c_2\,\rho}(H)\,=\,-\frac{g_A}{f_\pi}\left(\frac{g_\rho}{2}\right)^2 \vamn\,\hat \Gamma^{5\,m}_{1\,\mu}\,
\Delta^{\,\,\mu \nu}_\rho(q_2)\,\hat{\Gamma}^n_{2\,\nu}(q_2)\,+\,\ot\,.  \label{Mac2rp}  
\ee
The first contact term appears due to the piece $\sim$ $\vec {\rho}^{\,\mu}\times \vec {\Pi}$ in the current
(\ref{JAHLS}), while the second one comes from the second term on the right hand side of the Lagrangian
(\ref{LhlsNPRA1O}). 
It also holds that
\be
\hat{J}^{a\,\mu}_{a_1\,\rho}(a_1,H)\,=\,\hat{J}^{a\,\mu}_{a_1\,\rho}(a_1)\,,  \label{Jaa1ra1h}
\ee
with the current $\hat{J}^{a\,\mu}_{a_1\,\rho}(a_1)$ given in Eq.\,(\ref{Jaa1ra1}).
An additional contact current is now generated from the vertex 
$\sim$ $\vec {a}^{\,\nu} \times \vec {\rho}^{\,\mu}_{\,\,\nu}$ in the current (\ref{JAHLS})
\be
\hat{J}^{a\,\mu}_{a_1\,\rho}(H)\,=\,i g_A \frac{g^2_\rho}{2} \vamn \Delta^{\nu \lambda}_{a_1}(q_1)
\hat {\Gamma}^{5\,m}_{1\,\lambda}\,\left[q^\mu_2\,\Delta^{\rho\,\,\,\eta}_{\,\,\nu}(q_2) - 
q_{2 \nu} \Delta^{\,\mu\,\eta}_{\rho}(q_2)\right]\,\hat{\Gamma}^n_{2\,\eta}(q_2)\,+\,\ot\,.  \label{Jaa1rH}
\ee
The last current we consider is the mesonic current
\be
\hat{J}^{a\,\mu}_{a_1\,\rho}(\pi,H)\,=\,i f_\pi q^\mu\,\Delta^\pi_F(q^2)\,
\hat{\cal M}^a_{a_1\,\rho}(H)\,,  \label{Jaa1rpH}
\ee
with
\bea
\hat{\cal M}^a_{a_1\,\rho}(H)\,&=&\,-\frac{g_A}{f_\pi} \left( \frac{g_\rho}{2} \right) ^2\, \vamn
\hat \Gamma^{5\,m}_{1\,\lambda}\,\left[\,(2q_{2\,\zeta} q_{2\,\nu} + q_{2\,\zeta} q_{1\,\nu} - q_{1\,\zeta} q_{1\,\nu})
\,\Delta^{\,\zeta\,\lambda}_{a_1}(q_1)  \right. \nonumber \\
& &  \left. +\,
(-2q^2_2 - q_1 \cdot q_2 + q^2_1)\,\Delta^{\,\nu\,\lambda}_{a_1}(q_1)\,\right] 
\, \Delta^{\,\nu \lambda}_{\rho}(q_2)\,
\hat{\Gamma}^n_{2\,\eta}(q_2)\,+\,\ot\,.   \label{Maa1rpH}
\eea
It is strightforward to find that the the $\rho$ meson axial MECs of this section 
\be
\hat{\tilde J}^{a\,\mu}_{BS\,\rho}(ex,H)\,=\,\hat{J}^{a\,\mu}_{c_1\,\rho}(\pi,H) + \hat{J}^{a\,\mu}_{c_2\,\rho}(\pi,H) +
\hat{J}^{a\,\mu}_{a_1\,\rho}(a_1,H) + \hat{J}^{a\,\mu}_{a_1\,\rho}(H)
+ \hat{J}^{a\,\mu}_{a_1\,\rho}(\pi,H)\,,  \label{JTREXHBS}
\ee
satisfies the WTI
\be
q_\mu \hat{\tilde J}^{a\,\mu}_{BS\,\rho}(ex,H)= \left[\hat{e}_A(1)+\hat{e}_A(2),\hat{V}_\rho \right]_+
+i f_\pi m^2_\pi\,\Delta^\pi_F(q^2)\hat{\tilde  {\cal M}}^a_\rho(2,H)\,,
  \label{dJTREXHBS}
\ee
with the pion absorption amplitude given by
\be
\hat {\tilde {\cal M}}^a_\rho(2,H)\,=\,\hat{\cal M}^a_{c_1\,\rho}(H)\,+\,\hat{\cal M}^a_{c_2\,\rho}(H)\,+\,
\hat{\cal M}^a_{a_1\,\rho}(H)\,. 
\label{Mtar2H}
\ee

As to the $a_1$ meson axial MECs, we have now only the current 
\be
\hat{\tilde J}^{a\,\mu}_{BS\,a_1}(ex,H)\,=\,\hat {J}^{a\,\mu}_{c_1\,a_1}(\pi)\,,
\label{JRA1EXHBS}
\ee
of the currents (\ref{Jacia1}), for which the WTI (\ref{dJac1a1}) is valid.

\subsection{The continuity equation for the current matrix element \label{CH33}}

In analogy with Eq.\,(\ref{JBSE}), the full BS weak axial current is now 
\be
\hat{J}^{a\,\mu}_{BS}(H)\,=\,\hat{J}^{a\,\mu}_{IA}\,+\,\hat{J}^{a\,\mu}_{BS}(ex,H)\,, \label{JBSEH}
\ee
where 
\bea
\hat{J}^{a\,\mu}_{BS}(ex,H)\,&=&\,\hat{J}^{a\,\mu}_{BS\,\pi}(ex,H)\,+\,\hat{J}^{a\,\mu}_{BS\,\rho}(ex,H)
\,+\,\hat{J}^{a\,\mu}_{BS\,a_1}(ex,H)\,+\,\hat{J}^{a\,\mu}_{c\,\omega}(\pi)\,,   \label{JBSEXH}  \\
\hat{J}^{a\,\mu}_{BS\,\pi}(ex,H)\,&=&\,\hat{J}^{a\,\mu}_{BS\,\pi}(p)\,+\,
\hat{J}^{a\,\mu}_{BS\,\pi}(np,H)\,+\,\hat{J}^{a\,\mu}_{\pi}(\Delta)\,,  \label{JBSPEH}   \\
\hat{J}^{a\,\mu}_{BS\,\rho}(ex,H)\,&=&\,\hat{\tilde J}^{a\,\mu}_{BS\,\rho}(ex,H)\,+\,
\hat{J}^{a\,\mu}_{\rho}(\Delta)\,, \label{JBSREH}  \\
\hat{J}^{a\,\mu}_{BS\,a_1}(ex,H)\,&=&\,\hat{\tilde J}^{a\,\mu}_{BS\,a_1}(ex,H)\,+\,
\hat{J}^{a\,\mu}_{a_1}(\Delta)\,.  \label{JBSA1EH}  
\eea
The associated pion absorption amplitudes read
\bea
\hat{\cal M}^a(H)\,&=&\,i\hat{\Gamma}^a_1 G^{-1}_2+i\hat{\Gamma}^a_2 G^{-1}_1+
\hat{\cal M}^a_\pi(2)+\hat{\cal M}^a_{\rho}(2,H)+\hat{\cal M}^a_{a_1}(2,H)
+\hat{\cal M}^a_{c\,\omega}\,,  \label{MaH}  \\
\hat{\cal M}^a_{\rho}(2,H)\,&=&\,\hat{\tilde {\cal M}}^a_{\rho}(2,H)\,+\,\hat{\cal M}^a_{\rho}(\Delta)\,,
\label{MarH}  \\
\hat{\cal M}^a_{a_1}(2,H)\,&=&\,\hat{\cal M}^a_{c_1\,a_1}\,+\,\hat{\cal M}^a_{a_1}(\Delta)\,. \label{Mara1H}
\eea
An analogue of Eq.\,(\ref{divf}) is then
\be
q_\mu \hat{J}^{a\,\mu}_{BS}(H)\,=\,[\,\hat{e}_A(1)+\hat{e}_A(2),\,{\cal G}^{-1}\,]_+\,
+\,if_\pi\,m^2_\pi\Delta^\pi_F(q^2)\,\hat{\cal M}^a(H)\,,  \label{divfH}
\ee
which again leads to the standard PCAC constraint
\be
q_\mu <\psi|\,\hat{J}^{a\,\mu}_{BS}(H)\,|\psi>\,=\,if_\pi\,m^2_\pi\Delta^\pi_F(q^2)\,
<\psi|\,\hat{\cal M}^a(H)\,|\psi>\,.  \label{divmeH}
\ee

The possible model dependence is given by effects due to the difference between 
the currents $\hat{J}^{a\,\mu}_{BS}(ex)$ and $\hat{J}^{a\,\mu}_{BS}(ex,H)$, which
reduces to investigating the difference between the currents
\be
\hat J^{a\,\mu}(M)\,=\,\hat{J}^{a\,\mu}_{c_1\,\pi}(a_1) + \hat{J}^{a\,\mu}_{\rho\,\pi}(a_1)
 + \hat{J}^{a\,\mu}_{a_1\,\rho}(\pi) +  \hat {J}^{a\,\mu}_{c_2\,a_1}(\pi)
+ \hat {J}^{a\,\mu}_{c_3\,a_1}(\pi)\,,
\label{JaM} 
\ee
and
\be
\hat J^{a\,\mu}(H)\,=\,\frac{1}{2}\hat{J}^{a\,\mu}_{\rho\,\pi}(1,H) + \hat{J}^{a\,\mu}_{\rho\,\pi}(2,H)
+ \hat{J}^{a\,\mu}_{\rho\,\pi}(a_1,H) + \hat{J}^{a\,\mu}_{a_1\,\rho}(H) + \hat{J}^{a\,\mu}_{a_1\,\rho}(\pi,H)
+ \hat{J}^{a\,\mu}_{c_2\,\rho}(\pi,H)\,.   \label{JaH} 
\ee
The presence of the $\pi$ meson axial MECs entering these two equations may indicate that the 
model dependence can appear already at relatively low energies.
However, inspecting the structure of the currents
$\hat{J}^{a\,\mu}_{c_1\,\pi}(a_1)$ and $\frac{1}{2}\hat{J}^{a\,\mu}_{\rho\,\pi}(1,H)$ shows that
the leading terms cancel each other and only the momentum dependent terms survive. 
So the model dependence should be expected
presumably of the short range nature and can be expected at high energies.


\section{Results and conclusions  \label{RC}}

In this paper, we have investigated the structure of the axial MECs for the
$N \Delta \pi \rho\,a_1 \omega$ system in the conjunction with the BS equation.
Our starting points are Lagrangians and one-body currents of this
system (Sect.\,\ref{CH1}). One of the Lagrangians, ${\cal L}^M$, is
the minimal one and it is approximately invariant
under the local $[SU(2)_L\times SU(2)_R]\times U(1)$ symmetry. The heavy meson fields
are introduced as massless Yang-Mills compensating fields belonging to the linear
realization of the chiral symmetry. The symmetry is violated by the heavy meson masses,
which are introduced, together with the external electroweak interaction, by hands.

Another choice considered here, is the Lagrangian ${\cal L}^H$, which reflects the
$[SU(2)_L\times SU(2)_R]_l \times U(1)_l$
hidden local symmetry.
In this case, the symmetry is not violated by the heavy meson mass terms and the external electroweak
interaction enters naturally into the scheme as gauge fields of the initial
global chiral group $[SU(2)_L\times SU(2)_R]_g$.
We take the heavy meson fields belonging to the non-linear realization of the HLS,
having in mind that the Lagrangian expressed in terms of the fields belonging to the
linear realization of the HLS is physically equivalent. Let us remind that the transition
between the representations is done using the Stueckelberg transformation.

Essentially, both Lagrangians differ by the choice of correction Lagrangians needed to
obtain correct amplitudes of elementary processes up to the energies $\sim$ 1 GeV.

In Sect.\,\ref{CH2},
(a) the $\pi$, $\rho$, $a_1$ and $\omega$ meson axial MECs  from the
minimal Lagrangian ${\cal L}^M$ are derived. These currents satisfy the WTIs separately but the case
of the $\rho$ and $a_1$ meson axial MECs, which should be considered
together,
(b) the WTI for the full BS current and the divergence of its matrix element between the
two-body BS wave functions is given and it is shown that it satisfies the standard PCAC
constraint and
(c) the introduction of the strong form factors into the BNN vertices is discussed.

The derivation of the analogous currents for the Lagrangian ${\cal L}^H$ is presented in
Sect.\,\ref{CH3}, where also the model dependence is indicated.

The derived currents, with the nucleon Born terms added and sandwiched between the Dirac
spinors, can be used in the standard nuclear physics calculations. To our knowledge,
more complete set of the axial MECs has not yet been published.

In comparison with earlier works \cite{Dm}, \cite{An}, we can say that the set of the axial MECs
derived here is more realistic. Indeed, besides the $a_1$ meson exchange,
the considered $\pi$, $\rho$ and $\omega$ meson exchanges enter the realistic NN potentials of the OBE type \cite{Bo}.
Let us note that the $a_1$ meson exchange was included into OBEPQB in \cite{OPT}.
From the meson exchanges which are still not included, only the $\sigma$ meson exchange is of importance
for OBEPs in describing the NN attraction at medium distances. This particle is controversial
and it is difficult to include it consistently. It enters the linear $\sigma$ model
naturally as a partner of the $\pi$ meson. However, the mass of the resonance \cite{PD}
having the quantum numbers of the $\sigma$ meson is by a factor $\sim$ 2 larger than its mass
extracted from the fit of OBEPs to the data \cite{Bo}. On the other hand, in the
non-linear $\sigma$ models, to which our models belong, this particle does not enter
by construction and can be put in by hands only.

Besides, Ref.\,\cite{Dm} contains an error \cite{Sm} in sections dealing with the non-linear $\sigma$
meson model. Indeed, it should be $(p'_1 - p_1 - 2q)_\nu$ in Eq.\,(40) instead of
$(p'_1 - p_1 - q)_\nu$, which removes (a) the confusion about the axial current nonconservation
even in the chiral limit given in the paragraph after this equation and
(b) the $q$ dependence on the right hand side of Eq.\,(60).
It is seen then that the correct current can be simply redefined so that it satisfies the WTI and
the matrix element of its divergence PCAC.

As the next step would be calculations of the cross sections for the weak reactions like
Eqs.\,(\ref{NCN})-(\ref{CCA}). In order to do them consistently, one has to use solutions
of the BS equation for the nuclear wave functions  using a OBEP constructed in the
same model. Such calculations would give more confidence in conclusions about the
charged and neutral processes in neutrino disintegration of the deuteron and consequently,
about the neutrino oscillations. These are important considerations for understanding
the solar neutrino experiments at SNO and
other possible experiments at higher energies.

\acknowledgments

This work is partially supported by the grant GA \v{C}R 202/97/0447. A part of this work was
done during the stays of E.\,T.\, at the TJNAL and at the
Department of Physics, University of
Alberta. He thanks Prof.\,F.\,L.\,Gross and Prof.\,F.\,C.\,Khanna
for the hospitality and Prof.\,F.\,L.\,Gross
and Mgr.~J.~Smejkal for discussions.
Research of F.~C.~K.~is supported in part by NSERCC.


\newpage
\hspace{130pt}
\begin{picture}(10000,15000)
\drawline\fermion[\N\REG](0,0)[13948]
\drawarrow[\N\ATBASE](\pmidx,\pmidy)
\drawarrow[\N\ATBASE](0,2000)
\drawarrow[\N\ATBASE](0,11948)
\put(2450,-1800){a}
\global\advance\pmidx by 450
\global\advance\pmidy by -300
\put(\pmidx,\pmidy){$P$}
\global\advance\pmidx by -2000
\put(\pmidx,\pmidy){$\cal N$}
\global\advance\fermionbackx by -1500
\put(\fermionbackx,\fermionbacky){$p_{\,1}^{\,\prime}$}
\put(\fermionbackx,0){$p_{\,1}$}
\global\advance\pmidy by -2974
\thicklines
\drawline\fermion[\W\REG](0,\pmidy)[2000]
\drawarrow[\E\ATBASE](\pmidx,\pmidy)
\global\advance\pmidx by -600
\global\advance\pmidy by 400
\put(\pmidx,\pmidy){$B$}
\thinlines
\drawline\photon[\W\REG](\fermionbackx,\fermionbacky)[4]
\global\advance\pmidx by -1600
\global\advance\pmidy by 800
\put(\pmidx,\pmidy){$\hat{J}^{a,\,\mu}(q)$}
\thicklines
\global\seglength=1400
\global\gaplength=350
\drawline\scalar[\E\REG](0,9948)[3]
\drawarrow[\E\ATBASE](\pmidx,\pmidy)
\global\advance\pmidx by -450
\thinlines
\global\advance\pmidy by 400
\put(\pmidx,\pmidy){$B_2$}
\global\advance\pmidy by -1300
\put(\pmidx,\pmidy){$q_{\,2}$}
\drawline\fermion[\N\REG](4900,0)[13948]
\drawarrow[\N\ATBASE](4900,2000)
\drawarrow[\N\ATBASE](4900,11948)
\global\advance\fermionbackx by 600
\put(\fermionbackx,\fermionbacky){$p_{\,2}^{\,\prime}$}
\put(\fermionbackx,0){$p_{\,2}$}
\end{picture}
\hspace{20pt}
\begin{picture}(10000,15000)
\drawline\fermion[\N\REG](0,0)[13948]
\drawarrow[\N\ATBASE](\pmidx,\pmidy)
\drawarrow[\N\ATBASE](0,2000)
\drawarrow[\N\ATBASE](0,11948)
\put(2450,-1800){b}
\global\advance\pmidx by 450
\global\advance\pmidy by -300
\put(\pmidx,\pmidy){$Q$}
\global\advance\pmidx by -2000
\put(\pmidx,\pmidy){$\cal N$}
\global\advance\fermionbackx by -1500
\put(\fermionbackx,\fermionbacky){$p_{\,1}^{\,\prime}$}
\put(\fermionbackx,0){$p_{\,1}$}
\global\advance\pmidy by 2974
\thicklines
\drawline\fermion[\W\REG](0,\pmidy)[2000]
\drawarrow[\E\ATBASE](\pmidx,\pmidy)
\global\advance\pmidx by -600
\global\advance\pmidy by 400
\put(\pmidx,\pmidy){$B$}
\thinlines
\drawline\photon[\W\REG](\fermionbackx,\fermionbacky)[4]
\global\advance\pmidx by -1600
\global\advance\pmidy by 800
\put(\pmidx,\pmidy){$\hat{J}^{a,\,\mu}(q)$}
\thicklines
\global\seglength=1400
\global\gaplength=350
\drawline\scalar[\E\REG](0,2974)[3]
\drawarrow[\E\ATBASE](\pmidx,\pmidy)
\global\advance\pmidx by -450
\thinlines
\global\advance\pmidy by 400
\put(\pmidx,\pmidy){$B_2$}
\global\advance\pmidy by -1300
\put(\pmidx,\pmidy){$q_{\,2}$}
\drawline\fermion[\N\REG](4900,0)[13948]
\drawarrow[\N\ATBASE](4900,2000)
\drawarrow[\N\ATBASE](4900,11948)
\global\advance\fermionbackx by 600
\put(\fermionbackx,\fermionbacky){$p_{\,2}^{\,\prime}$}
\put(\fermionbackx,0){$p_{\,2}$}
\end{picture}
\vspace{50pt}
\newline
\hspace{50pt}
\begin{picture}(10000,15000)
\drawline\fermion[\N\REG](0,0)[13948]
\drawarrow[\N\ATBASE](0,2000)
\drawarrow[\N\ATBASE](0,11948)
\put(2450,-1800){c}
\global\advance\fermionbackx by -1500
\put(\fermionbackx,\fermionbacky){$p_{\,2}^{\,\prime}$}
\put(\fermionbackx,0){$p_{\,2}$}
\global\advance\fermionbackx by 1500
\thicklines
\global\seglength=1400
\global\gaplength=350
\drawline\scalar[\E\REG](0,6974)[3]
\thinlines
\drawarrow[\W\ATBASE](\pmidx,\pmidy)
\global\advance\pmidx by -450
\global\advance\pmidy by 400
\put(\pmidx,\pmidy){$B_2$}
\global\advance\pmidy by -1300
\put(\pmidx,\pmidy){$q_{\,2}$}
\drawline\fermion[\N\REG](4900,0)[13948]
\drawarrow[\N\ATBASE](4900,2000)
\drawarrow[\N\ATBASE](4900,11948)
\global\advance\fermionbackx by 600
\put(\fermionbackx,\fermionbacky){$p_{\,1}^{\,\prime}$}
\put(\fermionbackx,0){$p_{\,1}$}
\thicklines
\drawline\fermion[\E\REG](\pmidx,\pmidy)[2000]
\drawarrow[\W\ATBASE](\pmidx,\pmidy)
\global\advance\pmidx by -600
\global\advance\pmidy by 400
\put(\pmidx,\pmidy){$B$}
\thinlines
\drawline\photon[\E\REG](\fermionbackx,\fermionbacky)[4]
\global\advance\pmidx by -1600
\global\advance\pmidy by 800
\put(\pmidx,\pmidy){$\hat{J}^{a,\,\mu}(q)$}
\end{picture}
\hspace{50pt}
\begin{picture}(10000,15000)
\drawline\fermion[\N\REG](0,0)[13948]
\drawarrow[\N\ATBASE](0,2000)
\drawarrow[\N\ATBASE](0,11948)
\put(4200,-1800){d}
\global\advance\fermionbackx by -1500
\put(\fermionbackx,\fermionbacky){$p_{\,1}^{\,\prime}$}
\put(\fermionbackx,0){$p_{\,1}$}
\thicklines
\global\seglength=1400
\global\gaplength=350
\drawline\scalar[\E\REG](0,9948)[5]
\thinlines
\drawarrow[\E\ATBASE](6000,\pmidy)
\drawarrow[\W\ATBASE](2400,\pmidy)
\put(1800,10450){$B_1$}
\put(6300,10450){$B_2$}
\put(1800,9000){$q_{\,1}$}
\put(6300,9000){$q_{\,2}$}
\drawline\photon[\S\REG](\pmidx,\pmidy)[7]
\global\advance\pmidx by 200
\global\advance\pmidy by -1100
\drawarrow[\N\ATBASE](\pmidx,\pmidy)
\global\advance\pmidx by -1600
\global\advance\pmidy by -3300
\put(\pmidx,\pmidy){$\hat{J}^{a,\,\mu}(q)$}
\drawline\fermion[\N\REG](8400,0)[13948]
\drawarrow[\N\ATBASE](8400,2000)
\drawarrow[\N\ATBASE](8400,11948)
\global\advance\fermionbackx by 600
\put(\fermionbackx,\fermionbacky){$p_{\,2}^{\,\prime}$}
\put(\fermionbackx,0){$p_{\,2}$}
\end{picture}
\hspace{50pt}
\begin{picture}(10000,15000)
\drawline\fermion[\N\REG](0,0)[13948]
\drawarrow[\N\ATBASE](0,2000)
\drawarrow[\N\ATBASE](0,11948)
\global\advance\fermionbackx by -1500
\put(\fermionbackx,\fermionbacky){$p_{\,1}^{\,\prime}$}
\put(\fermionbackx,0){$p_{\,1}$}
\put(4200,-1800){e}
\global\seglength=1400
\global\gaplength=350
\thicklines
\drawline\scalar[\E\REG](0,9948)[5]
\thinlines
\drawarrow[\E\ATBASE](6000,\pmidy)
\drawarrow[\W\ATBASE](2400,\pmidy)
\put(1800,10450){$B_1$}
\put(6300,10450){$B_2$}
\put(1800,9000){$q_{\,1}$}
\put(6300,9000){$q_{\,2}$}
\thicklines
\drawline\fermion[\S\REG](\pmidx,\pmidy)[2100]
\drawarrow[\N\ATBASE](\pmidx,\pmidy)
\global\advance\pmidx by 300
\global\advance\pmidy by -400
\put(\pmidx,\pmidy){$B$}
\thinlines
\drawline\photon[\S\REG](\fermionbackx,\fermionbacky)[5]
\global\advance\pmidx by -1600
\global\advance\pmidy by -3300
\put(\pmidx,\pmidy){$\hat{J}^{a,\,\mu}(q)$}
\drawline\fermion[\N\REG](8400,0)[13948]
\drawarrow[\N\ATBASE](8400,2000)
\drawarrow[\N\ATBASE](8400,11948)
\global\advance\fermionbackx by 600
\put(\fermionbackx,\fermionbacky){$p_{\,2}^{\,\prime}$}
\put(\fermionbackx,0){$p_{\,2}$}
\end{picture}

\vspace{10mm}
\newline
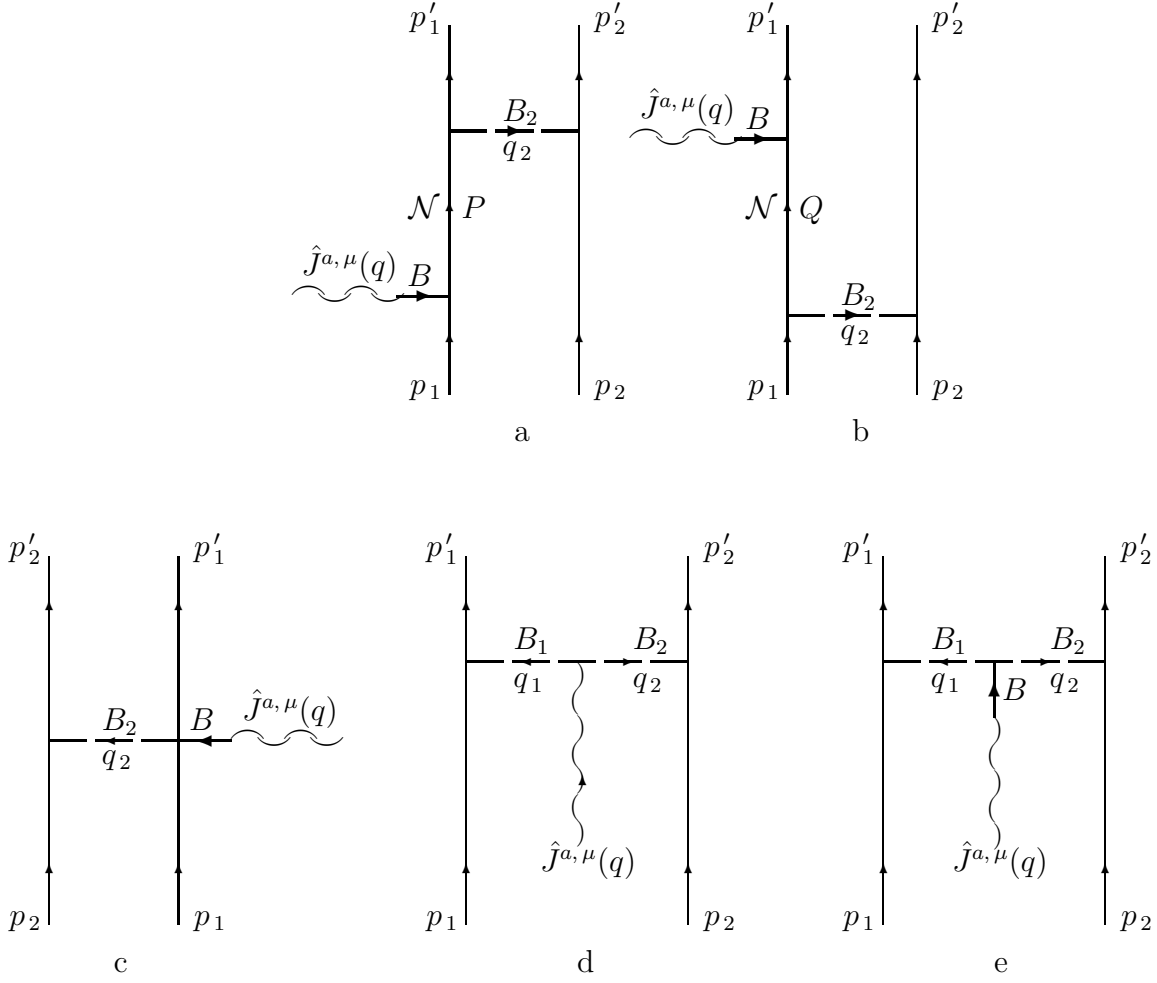
Fig.\,1. The general structure of the weak axial MEC operators considered in this paper.
The weak axial
interaction is mediated by the meson B which is either $\pi$ or $a_1$ meson.
The range of the
current is given by the meson $B_2$ which is here $\pi$, $\rho$, $a_1$ or $\omega$ meson.
The graphs a, b, represent the current ${\hat J}^{a\,\mu}_{B_2}({\cal N},\,B)$ with
$\cal N$
either for the nucleon N  or for the $\Delta(1236)$ isobar.
The graph c represents a contact current ${\hat J}^{a\,\mu}_{c\,B_2}(B)$.
It is connected with the weak
production amplitude of the $B_2$ meson on the nucleon. Another type of the contact
terms is given by the graph d, ${\hat J}^{a\,\mu}_{B_1\,B_2}$, where the weak axial
current interacts directly
with the mesons $B_1$ and $B_2$.
The graph e is for a mesonic current ${\hat J}^{a\,\mu}_{B_1\,B_2}(B)$.
The associated pion absorption amplitudes correspond to the graphs where
the weak axial interaction
is mediated by the pion, but with the weak interaction wavy line removed.
There are three types of these amplitudes in our models:
$\hat {\cal M}^a_{B_2}(\cal N)$,
$\hat {\cal M}^a_{c\,B_2}$ and $\hat {\cal M}^a_{B_1\,B_2}$.


\begin{thebibliography}{99}
\bibitem{Zu} K.~Zuber, Phys.~Rep. 305 (1998) 295.
\bibitem{JB} J.~Bahcall, in Proceedings of the XVth International  Conference
on Few-Body Problems in Physics, p.~29c, eds.~J.~C.~S.~Bacelar {\it et al}.,
North-Holland, 1998.
\bibitem{GK} K.~Grotz and H.~V.~Klapdor, The Weak Interaction in Nuclear,
Particle and Astrophysics, IOP Publishing Ltd, Bristol, England, 1990.
\bibitem{Rl} S.~P.~Riley {\it et al.}, Phys.~Rev. C 59 (1999) 1780.
\bibitem{Re} F.~Reines {\it et al.}, Phys.~Rev.~Lett. 45 (1980) 1307.
\bibitem{Vi} G.~S.~Vidiakin {\it et al.}, JETP Lett. 51 (1990) 245; \\
A.~G.~Vershinsky {\it et al.}, JETP Lett. 51 (1990) 82.
\bibitem{FV} P.~H.~Frampton and P.~Vogel, Phys.~Rep. 82 (1982) 339.
\bibitem{TK1} E.~Truhl\'{\i}k and F.~C.~Khanna, Few-Body Systems 11 (1991) 25.
\bibitem{SNO} The SNO Collaboration, The Sadbury Neutrino Observatory,
nucl-ex/9910016.
\bibitem{Ku} K.~Kubodera and S.~Nozawa, Int.~J.~Mod.~Phys. E 3 (1994) 101.
\bibitem{RP} R.~V.~Reid, Ann.~Phys.~(N.~Y.~) 50 (1968) 411; \\
M.~Lacombe {\it et al.}, Phys.~Rev. C 21 (1980) 861.
\bibitem{AVOG} J. Adam, Jr.\,, J.~W.~Van Orden and Franz Gross, Nucl.~Phys.
A 640 (1998) 391.
\bibitem{BS} E.~E.~Salpeter and H.~A.~Bethe, Phys.~Rev. 84 (1951) 1232.
\bibitem{KDR} K.~Kubodera, J.~Delorme and M.~Rho, Phys.~Rev.~Lett. 40 (1978) 755.
\bibitem{ITO} E.~Ivanov and E.~Truhl\'{\i}k, Sov.~J.~Part.~Nucl. 12 (1981) 198.
\bibitem{KT} M.~Kirchbach and E. Truhl\'{\i}k, Sov.~J.~Part.~Nucl. 17 (1986) 93.
\bibitem{To} I.~S.~Towner, Ann.~Rep.~Prog.~Nucl.~Part.~Sci. 36 (1986) 115; \\
I.~S.~Towner, Phys.~Rep. 155 (1987) 263.
\bibitem{Ma} J.-F.~Mathiot, Phys.~Rep. 173 (1989) 63.
\bibitem{Ri} D.~O.~Riska, Phys.~Rep. 181 (1989) 207.
\bibitem{Vo} V.~V.~Vorobyov {\it et al.}, Hyperfine Interactions 101/102 (1996) 413.
\bibitem{CT} J.~G.~Congleton and E.~Truhl\'{\i}k, Phys.~Rev. C 53 (1996) 956.
\bibitem{TS} E.~Truhl\'{\i}k and K.-M.~Schmitt, Few-Body Systems 11 (1992) 155.
\bibitem{ACo} F.~Ritz {\it et al.}, Phys.~Rev. C 55 (1997) 2214.
\bibitem{Sa} J.~J.~Sakurai, Currents and Mesons, The University of Chicago Press, 1968.
\bibitem{YM} C.~N.~Yang and R.~L.~Mills, Phys.~Rev. 96 (1954) 191.
\bibitem{BKY} M.~Bando, T.~Kugo and K.~Yamawaki, Phys.~Rep. 164 (1988) 217.
\bibitem{M} U.~-G.~Meissner, Phys.~Rep. 161 (1988) 213.
\bibitem{KM} N.~Kaiser and U.~-G.~Meissner, Nucl.~Phys. A 519 (1990) 671.
\bibitem{St} E.~G.~C.~Stueckelberg, Helv.~Phys.~Acta 14 (1941) 51.
\bibitem{IT1} E.~Ivanov and E.~Truhl\'{\i}k, Nucl.~Phys. A 316 (1979) 437.
\bibitem{TSA} E.~Truhl\'{\i}k, Czech.~J.~Phys. B 43 (1993) 467.
\bibitem{STG} J.~Smejkal, E.~Truhl\'{\i}k and H.~G\"oller, Nucl.~Phys. A
624 (1997) 655.
\bibitem{TK2} E.~Truhl\'{\i}k and F.~C.~Khanna, Int.~J.~Mod.~Phys. A 10 (1995) 499.
\bibitem{ATA} J.~Adam, Jr., E.~Truhl\'{\i}k and D.~Adamov\'a, Nucl.~Phys. A 494 (1989) 556.
\bibitem{GR} Franz Gross and D.~O.~Riska, Phys.~Rev. C 36 (1987) 1928.
\bibitem{W1} S.~Weinberg, The Quantum Theory of Fields II, Cambridge University Press, 1996.
\bibitem{STK1} J.~Smejkal, E.~Truhl\'{\i}k and F.~C.~Khanna, in Proc.~of the 7th Conf.~Mesons and
Light Nuclei '98, p.~490, eds.~J.~Adam {\it et al}., World Scientific, 1999.
\bibitem{Dm} V.~Dmitra\v{s}inovi\'c, Phys.~Rev. C 54 (1996) 3247.
\bibitem{An} S.~A.~Ananyan, Phys.~Rev. C 57 (1998) 2669.
\bibitem{OZ} V.~I.~Ogievetsky and B.~M.~Zupnik, Nucl.~Phys. B 24 (1970) 612.
\bibitem{STK2} J.~Smejkal, E.~Truhl\'{\i}k and F.~C.~Khanna, Few--Body Systems 26 (1999) 175.
\bibitem{BW} G.~E.~Brown and W.~Weise, Phys.~Rep. 22 (1975) 279.
\bibitem{Bo} R.~Machleidt, Adv.~Nucl.~Phys. 19 (1989) 189.
\bibitem{OPT} P.~Obersteiner, W.~Plessas and E.~Truhl\'{\i}k, in Proc.~of the 13th Conf.~
Particles and Nuclei, p.~430, ed.~A.~Pascolini, World Scientific, 1994.
\bibitem{PD} Particle Data Group, Phys.~Rev. D 54 (1996) 1.
\bibitem{Sm} J.~Smejkal, privat communication.


\end{thebibliography}
\end{document}